\documentclass[a4paper,11pt]{article}
\pdfoutput=1 
 
\usepackage{jcappub} 
 
\usepackage{aas_macros}
\usepackage{amssymb,amsmath}
\usepackage{graphicx}
\usepackage[dvipsnames]{xcolor}
\hypersetup{colorlinks=true,allcolors=teal}

\newcommand{\be}{\begin{equation}}
\newcommand{\ee}{\end{equation}}
\def\bear#1\ear{\begin{align}#1\end{align}}
\newcommand{\nline}{\notag \\}
\newcommand{\f}{\frac}
\newcommand{\de}{\mathrm{d}}
\newcommand{\del}{\partial}

\newcommand{\e}{\mathrm{e}}

\renewcommand{\mathbf}[1]{\mbox{\boldmath $#1$}}

\newcommand{\Msun}{\mathrm{M}_{\odot}}
\newcommand{\hcMpc}{h^{-1} \mathrm{cMpc}}

\newcommand{\eqn}[1]{eq.~(\ref{#1})}
\newcommand{\eqns}[2]{eqs.~(\ref{#1}) and (\ref{#2})}
\newcommand{\eqnsthree}[3]{eqs.~(\ref{#1}), (\ref{#2}) and (\ref{#3})}
\newcommand{\secn}[1]{section~\ref{#1}}
\newcommand{\appndx}[1]{appendix~\ref{#1}}
\newcommand{\fig}[1]{figure~\ref{#1}}

\newcommand{\Fig}[1]{Figure~\ref{#1}}
\newcommand{\tab}[1]{table~\ref{#1}}
\newcommand{\Tab}[1]{Table~\ref{#1}}

\usepackage[normalem]{ulem}  

\renewcommand{\citet}{\cite}

\title{Capturing Small-Scale Reionization Physics: A Sub-Grid Model for Photon Sinks with SCRIPT}

\author[a, 1]{Tirthankar Roy Choudhury,\note{Corresponding author.}}
\author[a]{Anirban Chakraborty}

\affiliation[a]{National Centre for Radio Astrophysics, TIFR,\\Pune University Campus, Ganeshkhind, Pune 411007, India}
 
\emailAdd{tirth@ncra.tifr.res.in}
\emailAdd{anirban@ncra.tifr.res.in}

\abstract{
    The epoch of reionization represents a major phase transition in cosmic history, during which the first luminous sources ionized the intergalactic medium (IGM). 
    However, the small-scale physics governing ionizing photon sinks—particularly the interplay between recombinations, photon propagation, and self-shielded regions—remains poorly understood. 
    Accurately modeling these processes requires a framework that self-consistently links ionizing emissivity, the clumping factor, mean free path, and photoionization rate. 
    In this work, we extend the photon-conserving semi-numerical framework, \texttt{SCRIPT}, by introducing a \emph{self-consistent sub-grid model} that dynamically connects these quantities to the underlying density field, enabling a more realistic treatment of inhomogeneous recombinations and photon sinks.
    We validate our model against a comprehensive set of observational constraints, including the UV luminosity function from HST and JWST, CMB optical depth from Planck, and Lyman-$\alpha$ forest measurements of the IGM temperature, photoionization rate, and mean free path. 
    Our fiducial model also successfully reproduces Lyman-$\alpha$ opacity fluctuations, reinforcing its ability to capture large-scale inhomogeneities in the reionization process. 
    Notably, we demonstrate that traditionally independent parameters, such as the clumping factor and mean free path, are strongly correlated, with implications for the timing, morphology, and thermal evolution of reionization.
    Looking ahead, we will extend this framework to include machine learning-based parameter inference. 
    With upcoming 21\,cm experiments poised to provide unprecedented insights, \texttt{SCRIPT} offers a powerful computational tool for interpreting high-redshift observations and refining our understanding of the last major phase transition in the universe.

}


\begin{document}

\date{}

\maketitle

\flushbottom

\section{Introduction}

Understanding the epoch of reionization is essential for reconstructing the formation and evolution of the first luminous sources and their impact on the intergalactic medium (IGM)~\cite{2001PhR...349..125B,2006PhR...433..181F,2016ARA&A..54..313M,2018PhR...780....1D}. While significant progress has been made in modeling reionization \cite{2022LRCA....8....3G,2022GReGr..54..102C}, a persistent challenge lies in accurately capturing the small-scale physics that govern ionizing photon sinks, particularly inhomogeneous recombinations, self-shielded regions, and fluctuations in the photoionization background. These processes play a crucial role in shaping the ionization history and structure of the IGM but remain difficult to resolve in large-scale simulations.

Recent advancements in observational capabilities have significantly enhanced our ability to probe reionization across multiple tracers. On one hand, the integrated reionization history is inferred from CMB anisotropy measurements \cite{2020A&A...641A...6P}, which probe the ionized component of the intergalactic medium (IGM) and indicate a reionization midpoint at $z \sim 7$. Moreover, measurements of the kinematic Sunyaev-Zel'dovich (kSZ) effect in the CMB temperature anisotropies \cite{2021ApJ...908..199R} provide constraints on the duration of reionization as well as insights into its potential sources \cite{2021MNRAS.501L...7C,2023MNRAS.522.2901J,2023MNRAS.526.3170N,2024MNRAS.530...35J}.

On the other hand, the later stages of reionization are investigated through Lyman-$\alpha$ (Ly$\alpha$) absorption spectra of quasars at $z \gtrsim 5$, particularly via analyses of Ly$\alpha$ opacity fluctuations \cite{2018MNRAS.479.1055B,2022MNRAS.514...55B}. In addition, observations of Ly$\alpha$ damping wing in the vicinity of high-redshift quasars \cite{2018ApJ...864..142D,2022MNRAS.512.5390G,2024ApJ...969..162D} and galaxies \cite{2024ApJ...971..124U} observed using the JWST has helped place independent constraints on the timing and progress of reionization. Beyond characterizing the ionized and neutral components of the IGM, the advent of the JWST has provided measurements of the ultraviolet luminosity functions (UVLFs) at high redshifts \cite{2023MNRAS.518.6011D,2023ApJS..265....5H,2023MNRAS.523.1009B,2024MNRAS.527.5004M,2024MNRAS.533.3222D} and offered valuable insights into the ionizing properties of galaxies \cite{2023NatAs...7..622C,2023MNRAS.524.2312E,2023A&A...672A.155M,2023A&A...672A.186P,2023ApJ...950...67M,2024MNRAS.527.4173L,2024ApJ...973....8H,2024Natur.626..975A,2024ApJ...975..245C,2024MNRAS.527.6139S,2024MNRAS.535.1067M,2025ApJ...981..134P}. Furthermore, measurements of the IGM temperature \cite{2019ApJ...872...13W,2020MNRAS.494.5091G} and the evolving mean free path derived \cite{2021MNRAS.508.1853B,2023ApJ...955..115Z} from quasar spectra offer additional windows into the physical processes governing reionization.

Constraining reionization robustly requires a combination of multiple observational probes, as relying on a single dataset can lead to significant degeneracies in reionization history. Comparing theoretical models with such a diverse range of observational data necessitates approaches that balance computational efficiency with the ability to capture sub-grid recombinations and photon propagation -- criteria that are well met by semi-numerical models of reionization.

Various theoretical frameworks have been employed in the literature to model cosmic reionization, each playing a vital role in accurately interpreting observational data. These methods include detailed, fully coupled radiation-hydrodynamical simulations \cite{2014ApJ...793...29G,2017MNRAS.472.4508S,2018MNRAS.479..994R,2020MNRAS.496.4087O,2022MNRAS.511.4005K}, post-processing of $N$-body simulations with radiative transfer calculations \cite{2006MNRAS.369.1625I,2008MNRAS.387..295A,2018MNRAS.476.1174E,2019MNRAS.485L..24K,2020MNRAS.491.1736K,2024A&C....4800861H}, and more recently developed computationally efficient semi-numerical simulations.  These semi-numerical models simplify radiative transfer into photon-counting algorithms and typically utilize coarser spatial resolutions compared to full radiative transfer simulations \cite{2004ApJ...613....1F,2009MNRAS.394..960C,2007ApJ...669..663M,2011MNRAS.411..955M,2010MNRAS.406.2421S,2017MNRAS.464.2992M,2018MNRAS.477.1549H,2018MNRAS.481.3821C,2023MNRAS.526.2942S}. Additionally, simpler analytical models have also been widely used for understanding the global evolution of reionization and gain insights into the average properties of the sources driving the process \cite{2003ApJ...586..693W, 2011MNRAS.413.1569M, 2012MNRAS.419.1480M,2013MNRAS.428L...1M, 2016MNRAS.460..417S, 2024ApJ...961...50S, 2024JCAP...07..078C}.

In recent years, we have developed an explicitly photon-conserving semi-numerical model of reionization,  \textbf{S}emi-numerical \textbf{C}ode for \textbf{R}e\textbf{I}onization with \textbf{P}ho\textbf{T}on-conservation (\texttt{SCRIPT})\footnote{\url{https://bitbucket.org/rctirthankar/script/}} \cite{2018MNRAS.481.3821C}, which is capable of computing a wide variety of observables. In its most basic form, the model generates an ionization field at a given redshift and has been employed to compare with CMB observations \cite{2021MNRAS.501L...7C,2023MNRAS.522.2901J,2024MNRAS.530...35J} as well as to forecast upcoming CMB polarization signals \cite{2021MNRAS.500..232P,2023MNRAS.522.2901J,2024MNRAS.527.2560J,2024MNRAS.530...35J} and the 21\,cm signal \cite{2023MNRAS.521.4140M,2024JCAP...03..027C}. The model has been further extended to incorporate inhomogeneous recombinations and to compute the thermal history, thereby enabling the self-consistent inclusion of radiative feedback effects \cite{2022MNRAS.511.2239M,2022MNRAS.515..617M,2023MNRAS.526.3920M}. This enhanced model has been compared with observations of the UV luminosity functions and the thermal properties of the IGM. Additionally, in a different work, we have implemented calculations of the photoionization rate using simple models for the ionizing mean free path, which can be employed to generate Ly$\alpha$ spectra for comparison with observations \cite{2021MNRAS.501.5782C}.

Despite these successes, there remain several avenues for improvement. For instance, in our treatment of the photoionization rate \cite{2021MNRAS.501.5782C}, a constant ionizing mean free path was assumed within ionized regions, whereas in reality, fluctuations are expected \cite{2016MNRAS.460.1328D}. Similarly, when modeling inhomogeneous recombinations, a simple parameterization of the clumping factor, the quantity which is the ratio of the number of recombinations to the number computed assuming a homogeneous IGM, was adopted \cite{2022MNRAS.511.2239M}; however, both the clumping factor and the mean free path are influenced by the distribution of self-shielded regions \cite{2000ApJ...530....1M,2005MNRAS.363.1031F} and hence their calculations must be inter-linked. Since such self-shielded regions, $\sim$ few kpc in size, cannot be resolved in semi-numerical simulations, we need to rely on physical sub-grid modeling to predict observables in a self-consistent manner \cite{2013ApJ...763..146E,2014MNRAS.440.1662S,2016ApJ...831...86P,2021MNRAS.506.2390Q, 2018MNRAS.477.1549H, 2020MNRAS.491.1600M, 2020ApJ...898..149D, 2021MNRAS.504.2443B, 2021ApJ...917L..37C,2024MNRAS.528.1296C}.

Thus building on the existing foundation, we extend \texttt{SCRIPT} to incorporate a self-consistent sub-grid model that dynamically links the clumping factor, ionizing mean free path, and photoionization rate to the density field. This improvement enables a more realistic treatment of self-shielded regions and inhomogeneous recombinations, crucial for accurately modeling the reionization history and IGM evolution. Unlike previous implementations that assumed simplified parameterized forms for the mean free path and recombination rate, our approach naturally captures spatial fluctuations in these quantities -- not only reflecting the two-phase nature of the IGM (ionized and neutral) but also capturing variations within ionized regions due to self-shielded structures. Although analytical models of reionization have used the connection between self-shielded density, recombinations and the mean free path \cite{2005MNRAS.361..577C,2005MNRAS.363.1031F,2011MNRAS.413.1569M,2012MNRAS.419.1480M,2013MNRAS.428L...1M}, implementing this relation within a simulation framework allows us to study spatial fluctuations more directly. While radiative transfer simulations can, in principle, capture such fluctuations by tracking the ionization and thermal histories consistently \cite{2007ApJ...671....1T,2015ApJ...813L..38D,2018MNRAS.473..560D,2019MNRAS.485L..24K,2020MNRAS.494.3080N,2020MNRAS.491.1736K,2021ApJ...923..161N,2024JCAP...12..025C,2025MNRAS.539L..18A}, resolving the sinks of ionizing photons in a cosmological volume remains challenging. Consequently, there is significant scope for incorporating sub-grid modeling. In this context, our model endeavors to capture these effects in a computationally efficient semi-numerical framework, albeit with free parameters that ultimately require observational calibration.

The primary aim of this paper is to develop the formalism underlying the sub-grid model and to demonstrate its utility in computing various observables. We provide a detailed exposition of the model and examine the implications of its underlying assumptions. Furthermore, we investigate the impact of simulation resolution and volume on the derived results. A comprehensive parameter inference analysis is beyond the scope of this paper and will be pursued in future work.

The paper is organized as follows: in \secn{sec:model}, we present the theoretical framework, emphasizing the sub-grid modeling approach introduced in this work. In \secn{sec:model_comparision_and_dependence}, we compare the predictions from our model, assuming a fiducial parameter set, with various observational datasets, and investigate how these predictions vary with different model parameters. Finally, in \secn{sec:summary}, we summarize our key findings and outline potential avenues for future research based on this model. The cosmological parameters adopted throughout this study are $\Omega_m = 0.308$, $\Omega_{\Lambda} = 1 - \Omega_m$, $\Omega_b = 0.0482$, $h = 0.678$, $n_s = 0.961$, and $\sigma_8 = 0.829$ \citep{2016A&A...594A..13P}.

\section{Theoretical Model}
\label{sec:model}

To achieve a reliable understanding of the reionization history, it is essential to use a theoretically sound and computationally manageable modeling framework. In this section, we introduce the theoretical model adopted in our study, emphasizing the novel sub-grid modeling techniques specifically developed to better represent important physical processes that affect reionization.

\subsection{Ionization sources}
\label{subsec:ionization_sources}

The theoretical model of reionization employed in this work is based on the semi-numerical code named \textbf{S}emi-numerical \textbf{C}ode for \textbf{R}e\textbf{I}onization with \textbf{P}ho\textbf{T}on-conservation (\texttt{SCRIPT})\footnote{\url{https://bitbucket.org/rctirthankar/script/}} \cite{2018MNRAS.481.3821C}. The inputs to \texttt{SCRIPT} are the large-scale density field (and velocity field, if desired) on a uniform grid at the redshift(s) of interest. The halo mass function $\left. \de n / \de M_h \right|_i$ in a grid ``cell'' labelled $i$ is generated via a subgrid prescription following the method based on conditional ellipsoidal collapse \citet{2002MNRAS.329...61S}.

In this work, we use GADGET-2 \citep{2005MNRAS.364.1105S} plugins provided by the 2LPT density field generator MUSIC \cite{2011MNRAS.415.2101H} (\url{https://www-n.oca.eu/ohahn/MUSIC/}) to generate the input $N$-body fields. Our default simulation box is of length $L_\mathrm{box} = 256h^{-1}$~cMpc with $256^3$ particles. We generate $151$ output snapshots at a fixed scale factor interval between redshift $z = 5$ and $20$. The particle positions and velocities are smoothed at an appropriate scale using the Cloud-In-Cell (CIC) kernel to generate the fields on a grid. Our default grid cell size is $\Delta x = 4 h^{-1}$~cMpc, but we test other values of $\Delta x$ in \appndx{app:resolution} to ensure numerical convergence of our results.

Galaxy properties are assigned to dark matter halos using a previously developed semi-analytical model \cite{2024JCAP...07..078C,2025arXiv250307590C}. To summarize, we assign each halo a UV luminosity at the rest wavelength $1500$\AA\ as
\be
L_{\mathrm{UV}, \mathrm{HI}, i}(M_h) = f_\star(M_h, z) \left(\f{\Omega_b}{\Omega_m}\right) l_\mathrm{UV}~M_h,
\ee
where the subscripts $\left(\mathrm{HI}, i\right)$ indicate that this relation applies to neutral regions in the $i$th grid cell, $f_\star(M_h, z)$ is the star-forming efficiency, and $l_\mathrm{UV}$ is the specific luminosity (i.e., luminosity per unit stellar mass). We assume that only halos heavier than the atomic cooling threshold (i.e., those with virial temperatures $> 10^4$~K) can form stars \citep{2001PhR...349..125B,2013MNRAS.432.3340S}. Redefining the efficiency parameter, the relation becomes
\be
L_{\mathrm{UV}, \mathrm{HI}, i}(M_h) = \varepsilon_\star(M_h, z)~l_\mathrm{UV, fid}~M_h,
\ee
where
\be
\varepsilon_\star(M_h, z) \equiv f_\star(M_h, z) \left(\f{\Omega_b}{\Omega_m}\right) \f{l_\mathrm{UV}}{l_\mathrm{UV, fid}}.
\ee
We choose $l_\mathrm{UV, fid} = 8.66 \times 10^{19}~\mathrm{erg}~\mathrm{s}^{-1}~\mathrm{Hz}^{-1}~\Msun^{-1}$, corresponding to continuous star formation over a time-scale of $100$ Myr with a $0.1-100 \Msun$ Salpeter IMF and metallicity $Z = 0.001 = 0.05 Z_\odot$, calculated using STARBURST99 v7.0.11 \cite{1999ApJS..123....3L}.

Radiative feedback from photoheating in ionized regions impacts star formation in lighter halos residing in those regions, slowing the progress of reionization. For ionized regions, photoheating increases the Jeans mass which, in each cell, is given by \citep{2021MNRAS.503.3698H}
\begin{equation}\label{eq:M_J}
    M_{J,i} = \frac{3.13 \times 10^{10} h^{-1} \Msun}{\Omega_m^{1/2}~(1+z)^{3/2}~\sqrt{18\pi^2}} ~\mu^{-3/2}~ \left(\frac{T_{\mathrm{HII}, i}}{10^4\mathrm{K}}\right)^{3/2},
\end{equation}
where $\mu$ is the mean molecular weight (assumed to be $0.59$, appropriate for ionized hydrogen and singly ionized helium) and $T_{\mathrm{HII}, i}$ is the temperature of the ionized regions in the cell. We assume the feedback to act gradually such that the gas fraction that remains inside a halo of mass $M_h$ is given by \citep{2013MNRAS.432.3340S,2019MNRAS.482L..19C}
\begin{equation}\label{eq:fg_gradual}
    f_{g, i}(M_h) = 2^{-M_{J,i} / M_{h}} = \exp\left(-\f{M_{J,i}}{1.44 M_h}\right),
\end{equation}
where $M_{J,i}$ is the Jeans mass in the cell $i$ (where the halo is situated) as defined in \eqn{eq:M_J}. For halos at the critical threshold mass $M_{J,i}$, the retained gas fraction is $50\%$, gradually decreasing for lighter halos. The above equation implicitly assumes that the gas fraction within halos responds instantaneously to changes in the ionization state and temperature of the surrounding intergalactic medium. In reality, the gas content is determined not solely by the instantaneous conditions but by the integrated thermal history of the region, with the response occurring over the halo dynamical timescale \cite{2000ApJ...542..535G,2021MNRAS.503.3698H,2022MNRAS.511.2239M}. Nevertheless, in our previous work \cite{2022MNRAS.511.2239M}, we demonstrated that neglecting this delayed response introduces only minimal deviations in the relevant observables, and these can be effectively absorbed by modest adjustments of model parameters. The mass-luminosity relation in ionized regions is then modified to
\be
L_{\mathrm{UV}, \mathrm{HII}, i}(M_h) = f_{g, i}(M_h)~\varepsilon_\star(M_h, z)~l_\mathrm{UV, fid}~M_h.
\ee

The UV magnitude is defined in terms of the luminosity as
\bear
M_{\mathrm{UV}, \mathrm{HI}, i} &= -2.5 \log_{10} \left(\f{L_{\mathrm{UV}, \mathrm{HI}, i}}{\mathrm{erg}~\mathrm{s}^{-1}~\mathrm{Hz}^{-1}}\right) + 51.6,
\nline
M_{\mathrm{UV}, \mathrm{HII}, i} &= -2.5 \log_{10} \left(\f{L_{\mathrm{UV}, \mathrm{HII}, i}}{\mathrm{erg}~\mathrm{s}^{-1}~\mathrm{Hz}^{-1}}\right) + 51.6.
\ear
This assignment immediately leads to the computation of the luminosity function in the $i$th cell as
\be
\Phi_{i}(M_{\mathrm{UV}}) = (1 - x_{\mathrm{HII}, i}) \left. \f{\de n}{\de M_h} \right|_i~\left|\f{\de M_h}{\de M_{\mathrm{UV}, \mathrm{HI}, i}}\right|
+ x_{\mathrm{HII}, i} \left. \f{\de n}{\de M_h} \right|_i~\left|\f{\de M_h}{\de M_{\mathrm{UV}, \mathrm{HII}, i}}\right|,
\label{eq:Phi_UV_cell}
\ee
where $x_{\mathrm{HII}, i}$ is the ionized fraction in the cell $i$. The global luminosity function is simply the average over all the cells in the simulation box
\be
\Phi(M_{\mathrm{UV}}) = \left \langle \Phi_i(M_{\mathrm{UV}}) \right \rangle.
\label{eq:Phi_UV_global}
\ee

As shown in our earlier work \cite{2024JCAP...07..078C}, matching the UVLF observations across a redshift range of $5 \lesssim z \lesssim 13$ requires the efficiency parameter to evolve non-trivially. We assume it follows a power-law dependence on halo mass:
\be
\varepsilon_\star(M_h, z) = \varepsilon_{\star, 10}(z) \left(\f{M_h}{10^{10}~\Msun}\right)^{\beta_\star(z)},
\label{eq:epsilon_star}
\ee
with both the normalization $\varepsilon_{\star, 10}(z)$ and the slope $\beta_\star(z)$ evolving with redshift as:
\bear
\log_{10} \varepsilon_{\star, 10}(z) &= \ell_{\star, 0} + \f{\ell_{\star, \mathrm{jump}}}{2} \tanh\left(\f{z - z_\mathrm{trans}}{\Delta z} \right),
\nline
\beta_\star(z) &= \beta_{\star, 0} + \f{\beta_{\star, \mathrm{jump}}}{2} \tanh\left(\f{z - z_\mathrm{trans}}{\Delta z} \right).
\label{eq:epsilon_star_tanh}
\ear
In this formulation, the parameter $\log_{10} \varepsilon_{\star, 10}$ asymptotes to $\ell_{\star, 0} - \ell_{\star, \mathrm{jump}} / 2$ at low redshifts and to $\ell_{\star, 0} + \ell_{\star, \mathrm{jump}} / 2$ at high redshifts, with the transition occurring at a characteristic redshift $z_\mathrm{trans}$ over a range $\Delta z$. The parameters for the slope $\beta_\star$ are interpreted similarly.

The next step in our analysis is to assign ionizing emissivities to galaxies. We relate the production rate of ionizing photons to the luminosity for halos in neutral regions as
\bear
\dot{N}_{\mathrm{ion}, \mathrm{HI}, i}(M_h) &= f_\mathrm{esc}(M_h, z)~\xi_\mathrm{ion}~L_{\mathrm{UV}, \mathrm{HI}, i}(M_h),
\nline
&= f_\mathrm{esc}(M_h, z)~\xi_\mathrm{ion}~\varepsilon_\star(M_h, z)~l_\mathrm{UV, fid}~M_h,
\ear
where $\xi_\mathrm{ion}$ is the rate of ionizing photons produced per unit UV luminosity at 1500 \AA\ and $f_\mathrm{esc}$ is the fraction of ionizing photons that escape into the IGM. This parameter $\xi_{\text{ion}}$ depends on the shape of the galaxy's ionizing spectrum, which in turn is influenced by properties such as the age, metallicity, and initial mass function of its stellar population. A similar relation holds for the production rate of ionizing photons in ionized regions, accounting for the effects of radiative feedback 
\be
\dot{N}_{\mathrm{ion}, \mathrm{HII}, i}(M_h) = f_\mathrm{esc}(M_h, z)~\xi_\mathrm{ion}~f_{g, i}(M_h)~\varepsilon_\star(M_h, z)~l_\mathrm{UV, fid}~M_h.
\ee
We define the escaping ionizing efficiency in terms of the escape fraction and $\xi_{\rm ion}$ as follows:
\be
\varepsilon_\mathrm{esc}(M_h, z) = f_\mathrm{esc}(M_h, z) \left(\f{\xi_\mathrm{ion}}{\xi_{\mathrm{ion}, \mathrm{fid}}}\right),
\ee
where we adopt the fiducial value $\xi_{\mathrm{ion}, \mathrm{fid}} = 10^{25.23}~\mathrm{erg}^{-1}~\mathrm{Hz}$. As in our previous works \cite{2024JCAP...07..078C}, we assume that the $M_h$-dependence of the escape fraction (and thus, the escaping ionizing efficiency) is given by
\be
\varepsilon_\mathrm{esc}(M_h, z) = \varepsilon_{\mathrm{esc}, 10} \left(\f{M_h}{10^{10}~\Msun}\right)^{\beta_\mathrm{esc}},
\label{eq:epsilon_esc}
\ee
where $\varepsilon_{\mathrm{esc}, 10}$ and $\beta_\mathrm{esc}$ are redshift-independent parameters. We thus assume that the escaping ionizing efficiency  for a given mass is constant across all redshifts, the reason being that with the present data it is very difficult to constrain the evolution. However, note that, the globally averaged escaping ionizing efficiency will evolve with redshift as the mass function evolves.

The production rate of ionizing photons for a halo is then given by:
\bear
\dot{N}_{\mathrm{ion}, \mathrm{HI}, i}(M_h) &= \varepsilon_\mathrm{esc}(M_h, z)~\varepsilon_\star(M_h, z)~\xi_{\mathrm{ion}, \mathrm{fid}}~l_{\mathrm{UV}, \mathrm{fid}}~M_h,
\nline
\dot{N}_{\mathrm{ion}, \mathrm{HII}, i}(M_h) &= f_{g,i}~\varepsilon_\mathrm{esc}(M_h, z)~\varepsilon_\star(M_h, z)~\xi_{\mathrm{ion}, \mathrm{fid}}~l_{\mathrm{UV}, \mathrm{fid}}~M_h.
\ear
The ionizing emissivity in the $i$th cell is given by:
\be
\dot{n}_{\mathrm{ion}, i} = \int_{M_\mathrm{cool}}^{\infty} \de M_h~\left. \f{\de n}{\de M_h} \right|_i~\left[(1 - x_{\mathrm{HII}, i})~\dot{N}_{\mathrm{ion}, \mathrm{HI}, i}(M_h) + x_{\mathrm{HII}, i}~\dot{N}_{\mathrm{ion}, \mathrm{HII}, i}(M_h) \right].
\ee
The integrated number of ionizing photons at a redshift $z$ is:
\be
n_{\mathrm{ion}, i}(z) = \int_\infty^z \de z'~\f{\de t'}{\de z'}~\dot{n}_{\mathrm{ion}, i}(z').
\ee
This quantity is used for generating the ionization fields using our photon-conserving algorithm.

\subsection{Ionization field}

The photon-conserving algorithm for generating ionization maps mainly consists of two steps. In the first step, we assign ionized regions of appropriate volumes around the ``source'' cells with $n_{\mathrm{ion}, i} > 0$. More specifically, we first consume $n_{H, i} + n_{\mathrm{rec}, i}$ number of these photons in the source cell itself, where $n_{\mathrm{rec}, i}$ is the integrated number of recombinations in the cell $i$ per unit comoving volume.. The remaining photons are then distributed to the other cells in increasing order of distance till all the photons from the source cell are exhausted. For a given cell $j$, if the number of photons available
\be
n_{\mathrm{ion, avail}, j} \geq n_{H, j}  + n_{\mathrm{rec}, j},
\ee the cell is flagged as completely ionized (and one is left with excess photons to be redistributed), else the cell is assigned an ionized fraction
\be
x_{\mathrm{HII}, j} = \f{n_{\mathrm{ion, avail}, j} - n_{\mathrm{rec}, j}}{n_{H,j}}.
\ee
This process is repeated independently for all source cells in the box. As a result, some of the grid cells which receive photons from multiple source cells may end up with $x_{\mathrm{HII}, j} > 1$ and are assigned as ``overionized''. In the second step, one distributes the excess ionizing photons in these unphysical overionized cells among the surrounding neighbouring cells which are yet to be fully ionized. The process is continued till all the overionized cells are properly accounted for. Clearly, the conservation of photon number is explicit in this model.

The comoving number density $n_{\mathrm{rec}, i}$ of recombinations for a cell can be computed by  solving a first order differential equation for the recombination rate density
\be
\label{eq:dnrec_dt}
\frac{\de n_{\mathrm{rec}, i}}{\de t} = \chi_{\mathrm{He}}~C_{H, i}~n^2_{H, i}~x_{\mathrm{HII}, i}~\alpha_A(T_{\mathrm{HII}, i})~(1 + z)^3,
\ee
where $\chi_{\mathrm{He}}$ is the contribution of singly-ionized helium to the free electron density, $C_{H, i} \equiv \left \langle n_{\mathrm{HII}}^2 \right \rangle_i / n_{H, i}^2$ is the clumping factor and $\alpha_A$ is the Case A recombination coefficient. While Case B recombination is typically appropriate for high-density, optically thick regions, where photons from recombinations are reabsorbed locally, it has been proposed that Case A recombination, when combined with an appropriately calibrated clumping factor, can also be employed to approximate ionization conditions in such environments \cite{2003ApJ...597...66M}. In our analysis, we adopt the Case A recombination coefficient as the default in order to facilitate direct comparison with results from other recent works \cite{2025MNRAS.539L..18A,2024arXiv241200799Q}, which commonly use this convention. However, we also explicitly evaluate the impact of using Case B recombination rates in \appndx{app:case_B}, in order to test the sensitivity of our results to this choice and ensure their robustness. The quantity relevant for generating the ionization maps is simply the integral
\be
n_{\mathrm{rec}, i} = \int_{\infty}^z \de z~\f{\de t}{\de z}~\f{\de n_{\mathrm{rec}, i}}{\de t}.
\ee

\subsection{Temperature of the IGM}
\label{sec:temp}

The implementation of radiative feedback and the computation of the recombination rate requires the temperature $T_i$ at each grid cell. The evolution of the kinetic temperature $T_i$ in a grid cell $i$ can be computed using the standard equation \cite{1997MNRAS.292...27H}
\begin{equation}
    \label{eq:dTdz}
    \frac{\de T_i}{\de z} = \f{2 T_i}{1 + z} + \f{2T}{3 \Delta_i} \f{\de \Delta_i}{\de z} + \f{2 \epsilon_i}{3 k_B n_{\mathrm{tot}, i}}\f{\de t}{\de z} + \frac{8 \sigma_T U n_{e, i}}{3 m_e c n_{\mathrm{tot}, i}} \left[T_\mathrm{CMB}(z) - T_i \right] \f{\de t}{\de z},
\end{equation}
where $\Delta_i$ is the cell overdensity, $\epsilon_i$ is the net heating rate per unit volume and $n_{\mathrm{tot},i}$ is comoving number density of all the gas particles (including free electrons). On the right hand, the first term corresponds to the cooling arising from the Hubble expansion, the second term is the adiabatic heating/cooling from structure formation, the third term gives the net heating from different astrophysical processes and the fourth term is the Compton cooling. The first two terms can be computed trivially as we already have the overdensity $\Delta_i$ at each redshift. For the third term, we include the photoheating from UV photons as this is the most dominant effect in the IGM at redshifts of our interest \citep{2016MNRAS.456...47M}. In the Compton cooling term, $U \propto T_\mathrm{CMB}^4(z)$ is radiation energy density and $T_\mathrm{CMB}(z) = 2.73~\mathrm{K}~(1+z)$ is the CMB temperature at redshfit $z$. Also, $\sigma_T$ is the Thomson scattering cross-section, $m_e$ the electron mass, $n_e$ the electron number density and $c$ the speed of light in vacuum.

The calculation of the photoheating term requires knowledge of the photoionization background in the ionized regions. In principle, since we compute the ionizing emissivities while generating the ionization maps, we should be able to compute the photoionizing background. This, however, requires knowledge of the mean free path and also introduces some further modeling challenges \cite{2021MNRAS.501.5782C}. A simpler way to implement the photoheating term is to assume photoionization equilibrium post-reionization and relate to the number of recombinations in the cell. The rate of change of temperature due to photoheating can be written as \cite{1997MNRAS.292...27H}
\begin{equation}
    \f{2 \epsilon_i}{3 k_B n_{\mathrm{tot}, i}} = \f{T_{\mathrm{re}}}{\chi_{\mathrm{He}}} \left[\chi_{\mathrm{He}}~C_{H, i}~n_{H,i}~x_{\mathrm{HII}, i}~\alpha_A(T_i)~(1 + z)^3 + \f{\de x_{\mathrm{HII}, i}}{\de t} \right],
    \label{eq:epsilon_PH}
\end{equation}
where $T_{\mathrm{re}}$ is the reionization temperature (i.e., the temperature of a region right after reionization). On the right hand side, the first term in the parentheses corresponds to heating in ionized regions post-reionizaton, while the second term is for the heating arising from newly ionized regions in the cells that are partially ionized.

Although our formalism is adequate for computing the average temperature of a cell (even when it is only partially ionized), we need the temperature of ionized regions within a partially ionized cell while implementing the radiative feedback. Now, temperatures of different ionized regions within a cell may be widely different as they get ionized at different times, hence one can only talk about an ``average'' temperature of ionized regions in our model. This average temperature of the ionized portion of a cell can be estimated as
\begin{equation}
    \label{eq:T_ion}
    T_{\mathrm{HII}, i} = \frac{T_i - (1-x_{\mathrm{HII}, i}) T_{\mathrm{HI}, i}}{x_{\mathrm{HII}, i}},
\end{equation}
where $T_{\mathrm{HI},i}$ is the temperature of neutral region in a cell which can be easily obtained from \eqn{eq:dTdz} by putting the photoheating and the Compton cooling terms to zero.

\subsection{Clumping factor}
\label{sec:clumping}

The calculation of the number of recombinations and also the heating rate requires knowledge of the clumping factor $C_{H,i}$ at each grid cell. The clumping factor of a region (or cell) of size $R$ and overdensity $\Delta_i$ is given by \citep{2022MNRAS.511.2239M}
\be
C_{H, i} = \f{\int_0^\infty \de \Delta~P_{V,i}(\Delta)~\Delta^2~x_\mathrm{HII}(\Delta)}{\Delta_i^2},
\ee
where $P_{V,i}(\Delta)$ is the conditional distribution of sub-grid overdensities $\Delta$ in the $i$th cell, $x_{\mathrm{HII}}(\Delta)$ is the ionized hydrogen fraction for the density element and we have ignored the mild temperature-dependence of $\alpha_A$ for simplicity. For low overdensities, we expect $x_{\mathrm{HI}}(\Delta) \ll 1$, while it approaches unity as $\Delta \approx \Delta_{\mathrm{ss}, i}$, the characteristic overdensity where the self-shielding becomes important. The exact form of $x_{\mathrm{HII}}(\Delta)$ and the value of $\Delta_{\mathrm{ss}, i}$ would depend on the photoionizing background in the cell. Usually these self-shielded regions are modeled using empirical fits from high-resolution simulations \citep{2013MNRAS.430.2427R,2018MNRAS.478.1065C}.

We consider a simple self-shielding model where $x_\mathrm{HII}(\Delta) = 1$ for $\Delta < \Delta_{\mathrm{ss}, i}$ and $0$ otherwise. Then
\be
C_{H, i} = \f{\int_0^{\Delta_{\mathrm{ss}, i}} \de \Delta~P_{V, i}(\Delta)~\Delta^2}{\Delta_i^2}.
\label{eq:C_H_i}
\ee
Although there exist more realistic models for computing the effect of self-shielding where the photoionization rate gradually reduces to zero at high densities, however, that requires knowledge of the density distribution. In our case, however, the simple model allows us to use simple scaling relations to model the sub-grid physics. We will discuss our modeling of $P_{V, i}(\Delta)$ in \secn{sec:conditional_pdf}.

The self-shielded threshold density $\Delta_{\mathrm{ss}, i}$ at each cell is calculated assuming that
\be
\label{eq:tau_for_deltaSS_cell}
N_\mathrm{HI}(\Delta_{\mathrm{ss},i})~\sigma_\mathrm{HI}(\nu_\mathrm{HI}) = 1,
\ee
where $N_\mathrm{HI}(\Delta)$ is the column density corresponding to a density element $\Delta$ and $\sigma_\mathrm{HI}$ is the photoionization cross-section. Under the assumption of dynamical equilibrium, we can relate the column density to the density using \citep{2001ApJ...559..507S}
\be
N_\mathrm{HI}(\Delta) = x_\mathrm{HI}(\Delta)~\bar{n}_H~\Delta~(1+ z)^3~L_J(\Delta),
\ee
where the Jeans length is given by
\be
L_J(\Delta) = \sqrt{\f{\gamma_c k_B (1 - Y)~(\Omega_b / \Omega_m)}{G~\mu~m_p^2}} ~ x_\mathrm{HI}(\Delta)~T^{1/2}(\Delta)~\Delta^{1/2}~\bar{n}_H^{1/2}~(1 + z)^{3/2},
\ee
with $\gamma_c$ being the ratio of specific heats, $T(\Delta)$ is the temperature at overdensity $\Delta$ and all other symbols have their usual meanings. We assume photoionization equilibrium at $\Delta \lesssim \Delta_{\mathrm{ss},i}$, so
\be
x_\mathrm{HI}(\Delta) = \f{\chi_\mathrm{He}~\alpha_A(T(\Delta))~\bar{n}_H~\Delta}{\Gamma_{\mathrm{HI}, i}} (1 + z)^3,
\ee
where we assume that the ionized regions within a given cell $i$ experience the same ionizing background characterized by the photoionization rate $\Gamma_{\mathrm{HI}, i}$.

Manipulating the above equations, along with the values $\gamma_c = 5/3$, $\Omega_b / \Omega_m = 0.16$, $1 - Y = 0.76$, $\mu = 0.59$, $\alpha_A(T) = 4.2 \times 10^{-13}~(T / 10^4~\mathrm{K})^{-0.7}~\text{cm}^3~\text{s}^{-1}$,  and $\sigma_\mathrm{HI}(\nu_\mathrm{HI}) = 6.3 \times 10^{-18}~\text{cm}^2$, we get
\be
\Delta_{\mathrm{ss},i} \approx 37 \chi_\mathrm{He}^{-2/3}
\left(\f{\bar{n}_H}{2 \times 10^{-7} \text{cm}^{-3}}\right)^{-1}
\left(\f{1+z}{8}\right)^{-3}
\left(\f{T_{\mathrm{HII}, i}}{10^4 \text{K}}\right)^{0.13}
\left(\f{\Gamma_{\mathrm{HI}, i}}{10^{-12}\text{s}^{-1}}\right)^{2/3},
\label{eq:Delta_ss}
\ee
where we have assumed $T(\Delta_{\mathrm{ss}, i}) = T_{\mathrm{HII}, i}$.\footnote{Some other studies \cite{2014MNRAS.440.1662S} obtain the $T$-dependence as $T^{0.17}$ because they assume the recombination rate $\alpha_A(T) \propto T^{-0.76}$ while we assume $\alpha_A(T) \propto T^{-0.7}$.}

Note that the above calculation of $\Delta_{\mathrm{ss},i}$ is accurate to only within factors $\mathcal{O}(1)$, as it neglects the spectral shape of the ionizing background when computing the photoionization cross-section in \eqn{eq:tau_for_deltaSS_cell}. Additionally, the exact numerical value of the Jeans length will depend on the geometry of the HI cloud.

The self-shielded threshold density thus depends on the photoionization rate $\Gamma_{\mathrm{HI}, i}$ in that cell.

\subsection{Photoionization rate}
\label{sec:photoionization}

The photoionization rate in a cell $j$ is given by
\be
\Gamma_{\mathrm{HI}, j} = (1 + z)^2 \f{\alpha_s}{\alpha_b + \alpha_\sigma} ~ \f{\sigma_\mathrm{HI}(\nu_\mathrm{HI})}{4 \pi} \sum_{i \neq j} \dot{N}_{\mathrm{ion}, i}~\f{\e^{-\tau_{i \to j}}}{x_{ij}^2},
\label{eq:Gamma_HI_j}
\ee
where $\dot{N}_{\textrm{ion}, j} \equiv \dot{n}_{\mathrm{ion}, j} (\Delta x)^3$ is the total number of ionizing photons produced per unit time in the cell $j$, $\tau_{i \to j}$ is the optical depth of ionizing photons between the cells $i$ and $j$, which is nothing but an integral along the line joining the two cells, and $x_{ij}$ is the comoving distance between the two cells. In the above expression, $\alpha_s$ is the spectral index of the ionizing sources, $\alpha_b$ is the spectral index of the ionizing background and $\alpha_\sigma$ is the spectral index of the hydrogen ionization cross-section. In this work, we use $\alpha_s=2.0$, $\alpha_b=1.2$, and $\alpha_\sigma=2.75$ \cite{2013MNRAS.436.1023B,2021MNRAS.508.1853B,2022MNRAS.511.2239M}.

We can rewrite the above expression as
\be
\Gamma_{\mathrm{HI}, j} = (1 + z)^2 \f{\alpha_s}{\alpha_b + \alpha_\sigma} ~ \f{\sigma_\mathrm{HI}(\nu_\mathrm{HI})}{4 \pi} (\Delta x)^3 \sum_{i \neq j} \gamma_{\mathrm{HI}}(i \to j),
\ee
where
\be
\gamma_{\mathrm{HI}}(i \to j) = \dot{n}_{\mathrm{ion}, i}~\f{\e^{-\tau_{i \to j}}}{x_{ij}^2},
\ee
is the contribution of sources in cell $i$ to the ionizing flux in cell $j$. The calculation of this quantity requires tracking the optical depth of all the cells that intersect the two cells $i$ and $j$, and thus can be computationally expensive. However, we can simplify the calculation by assuming statistical isotropy in the problem. Similar to what is done for generating the ionization fields, starting with the source cell $i$, we compute the average optical depth in spherical shells around the source and distribute the flux in all other cells $j$ in increasing order of cell distance from $i$. This reduces the computational requirement of the code, albeit at the expense of some accuracy. For example, an optically thick absorber in one direction may affect propagation of ionizing photons in other directions, which is clearly unphysical. However, because of statistical isotropy of the sources in the box, such effects are expected to be averaged out leading to a reasonable description of the $\Gamma_\mathrm{HI}$ field in the box. To assess the accuracy of our spherical averaging approach, we compare it against a simplified ray-tracing method in \appndx{app:ray-tracing}. We find that our method yields photoionization rate fields that are in good agreement with the ray-tracing results, particularly in astrophysical regimes relevant to this study -- such as at redshifts and ionization states where the mean free path of ionizing photons is relatively short.

Given the above simplification, we only need to assign the optical depth of ionizing photons in each grid cell. In the absence of neutral regions, it is simply
\be
\Delta \tau_i = \f{\Delta x}{\lambda_{\mathrm{ss}, i}},
\ee
where $\lambda_{\mathrm{ss}, i}$ is the mean free path of ionizing photons in the cell $i$, determined by the distance between self-shielded regions. We will discuss the calculation of $\lambda_{\mathrm{ss}, i}$ using the sub-grid conditional density distribution in the next section. In the presence of neutral regions, we assume that a fraction $1 - x_{\mathrm{HII}, i}$ of the ionizing photons absorbed are due to the neutral regions\footnote{
    Note that we are implicitly assuming that a volume fraction $1 - x_{\mathrm{HII}, i}$ is covered by neutral regions, while strictly speaking, the quantity is actually the mass-weighted neutral fraction. In case the neutral regions are in low-density regions, as would be the case for inside-out reionization, their volume fraction could be larger than what we assume. This is difficult to model without further assumptions, or without using a finer resolution grid. We will discuss the effect of this assumption when we check for convergence with respect to resolution in \appndx{app:resolution}.
},
while the rest are by the self-shielded regions in the ionized regions. We can then write the flux decrement as
\be
\e^{-\Delta \tau_i} = (1 - x_{\mathrm{HII}, i})~\e^{-\bar{n}_H \, \Delta_i \, \sigma_\mathrm{HI}(\nu_\mathrm{HI}) \, \Delta x \, (1 + z)^2} + x_{\mathrm{HII}, i}~\e^{-\Delta x / \lambda_{\mathrm{ss}, i}}.
\ee
For the typical grid sizes we use, the optical depth of the neutral regions $\gtrsim 100$, hence the first term on the right hand side is $\approx 0$. The optical depth of the cell then simplifies to
\be
\e^{-\Delta \tau_i} \approx x_{\mathrm{HII}, i}~\e^{-\Delta x / \lambda_{\mathrm{ss}, i}},
\ee
leading to an effective mean free path in the cell
\be
\lambda_{\mathrm{mfp}, i} \equiv \f{\Delta x}{\Delta \tau_i},
\label{eq:lambda_mfp_from_tau}
\ee
which in this case turns out to be
\be
\f{1}{\lambda_{\mathrm{mfp}, i}} \approx \f{1}{\lambda_{\mathrm{ss}, i}} + \f{1}{- \Delta x / \ln x_{\mathrm{HII}, i}}.
\ee

To complete the computation of the photoionization rate, we need to add the contribution of the source cell $i$ to $\Gamma_{\mathrm{HI}, i}$ in the same cell. Assuming that the sources within the grid cell are distributed uniformly, we can write the contribution as \cite{2016MNRAS.460.1328D,2021MNRAS.501.5782C}
\be
\Gamma_{\mathrm{HI}, i}^\mathrm{local} = x_{\mathrm{HII}, i} \,(1+z)^2 \f{\alpha_s}{\alpha_b + \alpha_\sigma} ~ \f{\sigma_\mathrm{HI}(\nu_\mathrm{HI})}{4 \pi} (\Delta x)^3 \,\dot{n}_{\mathrm{ion}, i} \left(1 - \e^{-r_0 / \lambda_{\mathrm{ss}, i}}\right) \f{3 \lambda_{\mathrm{ss}, i}}{4 \pi r_0^3},
\label{eq:Gamma_HI_local}
\ee
where $r_0 = \Delta x \, (3 / 4 \pi)^{1/3}$ is the radius of the sphere corresponding to the grid volume. The additional factor of $x_{\mathrm{HII}, i}$ is due to the fact that only ionized portions within the cell contribute to the ionization background.

\subsection{Mean free path of ionizing photons}

The calculation of $\Gamma_{\mathrm{HI}, i}$ requires knowledge of $\lambda_{\mathrm{ss}, i}$, the mean free path corresponding to the distance between self-shielded regions.

For calculating $\lambda_{\mathrm{ss}, i}$, we can assume that the cell is completely ionized so that we need not be concerned with neutral islands, their effect on the mean free path can be included through the value of $x_{\mathrm{HII}, i}$, as shown in the previous section.

As the photons travel through the IGM, the flux decreases due to absorption in HI absorbers \citep{1996ApJ...461...20H}. The corresponding effective optical depth can be related to the \emph{comoving} mean free path as
\be
\lambda_{\mathrm{ss}, i} = \f{c}{H_0 (1+z)^2} \f{1}{\int_0^{\infty} \de N_\mathrm{HI}~ f_i(N_\mathrm{HI}, z)~ \left(1 - \e^{-N_\mathrm{HI}~\sigma_\mathrm{HI}(\nu_\mathrm{HI})} \right)}
\ee
where
\be
f_i(N_\mathrm{HI}, z) \equiv \left. \f{\del^2 N}{\del N_\mathrm{HI} \del z} \right|_i~\f{H(z)}{H_0} \f{1}{(1+z)^2},
\ee
with $\del^2 N / \del N_\mathrm{HI} \del z|_i$ being the redshift and column density distribution of the HI absorbers in the $i$th grid cell. The above relations assume that the photons are absorbed at distances much shorter than the Hubble scale, valid till $z \sim 2$.

It is possible to connect $f_i(N_\mathrm{HI}, z)$ to $P_{V, i}(\Delta)$, e.g., by calculating the density of hydrogen using both the quantities. A straightforward calculation shows that \cite{2014MNRAS.440.1662S}
\be
\de N_\mathrm{HI}~f_i(N_\mathrm{HI}, z) = \f{c}{H_0}~\bar{n}_H~ \de \Delta~P_{V, i}(\Delta)~ \Delta~\f{x_\mathrm{HI}(\Delta)}{N_\mathrm{HI}(\Delta)}.
\ee
It then follows that the expression for the mean free path is
\be
\f{1}{\lambda_{\mathrm{ss}, i}} = \f{1}{\lambda_0} \int_0^{\infty} \de \Delta~P_{V, i}(\Delta)~ \Delta^{1/2}~\left(1 - \e^{-N_\mathrm{HI}(\Delta)~\sigma_\mathrm{HI}(\nu_\mathrm{HI})} \right) \left[\f{T(\Delta)}{10^4~\text{K}}\right]^{-1/2},
\ee
where
\be
\lambda_0 = \sqrt{\f{\gamma_c k_B \times 10^4 \text{K} \times (1 - Y)~(\Omega_b / \Omega_m)}{G~\mu~m_p^2}}~\bar{n}_H^{-1/2} ~(1 + z)^{-1/2}.
\ee
As before, using $\gamma_c = 5/3$, $1 - Y = 0.76$, $\Omega_b / \Omega_m = 0.16$, $\mu = 0.59$, we get
\be
\lambda_0 = 0.41~\text{Mpc} \left(\f{\bar{n}_H}{2 \times 10^{-7}~\text{cm}^{-3}}\right)^{-1/2} \left(\f{1+z}{8}\right)^{-1/2}.
\ee

Now as before, we assume $x_\mathrm{HII}(\Delta) = 1$ for $\Delta < \Delta_{\mathrm{ss}, i}$ and $0$ otherwise, i.e., $x_\mathrm{HI}(\Delta) = 1$ for $\Delta > \Delta_{\mathrm{ss}, i}$ and $0$ otherwise. Then $N_\mathrm{HI}(\Delta) = 0$ for $\Delta < \Delta_{\mathrm{ss}, i}$. For $\Delta > \Delta_{\mathrm{ss}, i}$, we assume $N_\mathrm{HI}(\Delta)~\sigma_\mathrm{HI}(\nu_\mathrm{HI}) \gg 1$ for the region to be self-shielded. In reality, there will be a smooth transition from optically thin to completely self-shielded regions, we have simplified the situation by assuming a sharp jump. With these assumptions, we get
\be
\f{1}{\lambda_{\mathrm{ss}, i}} = \f{1}{f_s \lambda_0} \int_{\Delta_{\mathrm{ss},i}}^{\infty} \de \Delta~P_{V, i}(\Delta)~ \Delta^{1/2} \left[\f{T(\Delta)}{10^4~\text{K}}\right]^{-1/2},
\label{eq:lambda_ss_i_P_V}
\ee
where $f_s$ is a $\mathcal{O}(1)$ numerical factor which corrects for the approximations made above.

\subsection{Conditional density distribution}
\label{sec:conditional_pdf}

As is clear from the above discussion, the conditional density distribution $P_{V, i}(\Delta)$ plays a central role in the calculation of the sub-grid quantities like the clumping factor and mean free path. In general, one expects this distribution to depend on the history of the spatial location under consideration. It is non-trivial to account for all such compexities in our model. We assume that $P_{V, i}(\Delta)$ depends only on the density $\Delta_i$ of the cell $i$ and the size of cell, i.e.,
\be
P_{V, i}(\Delta) = P_V(\Delta | \Delta_i; \Delta x).
\ee
The conditional distribution must satisfy the normalization conditions
\bear
\int_0^\infty \de \Delta \, P_V(\Delta | \Delta_i; \Delta x) &= 1, \nline
\int_0^\infty \de \Delta \, P_V(\Delta | \Delta_i; \Delta x) \, \Delta &= \Delta_i.
\ear
The unconditional distribution $P_V(\Delta)$ can be obtained by averaging over all the cells in the box, i.e.,
\be
P_V(\Delta) = \int_0^\infty \de \Delta_i \, P_V(\Delta | \Delta_i; \Delta x) \, P(\Delta_i),
\ee
where $P(\Delta_i)$ is the distribution of the cell overdensities in the box. Also note that as the size of the grid cell becomes large and its density approaches the mean density, the conditional distribution must approach the underlying unconditional distribution, i.e.,
\be
P_V(\Delta | \Delta_i = 1; \Delta x \to \infty) = P_V(\Delta).
\ee

There exist several forms for the unconditional density distribution $P_V(\Delta)$, either motivated by numerical simulations and physical considerations \cite{1997ApJ...479..523B,2000ApJ...530....1M,2001MNRAS.322..561C,2001ApJ...559...29C,2005MNRAS.361..577C,2015MNRAS.454L..76M} or fits to the distribution in hydrodynamical simulations \cite{2009MNRAS.394.1812P,2009MNRAS.398L..26B}. The conditional distribution is much more difficult to obtain from simulations as they require high dynamic range. Given these uncertainties, we attempt to model the sub-grid physics with as less assumptions regarding $P_V(\Delta | \Delta_i; \Delta x)$ as possible. From now on, we omit the explicit presence of grid size $\Delta x$ in the notation.

The first point to note is that for both $C_{H, i}$ and $\lambda_{\mathrm{ss}, i}$, the integrals are determined by the behaviour of $P_V(\Delta | \Delta_i)$ around $\Delta \sim \Delta_{\mathrm{ss}, i}$, the self-shielded threshold density. Also, since the self-shielded regions are expected to be in high-density regions, we only need to model the high-density behaviour of the conditional distribution \cite{2011ApJ...743...82M}. Now, the form of the unconditional PDF $P_V(\Delta)$ is such that it follows a power law distribution at $\Delta \sim \Delta_\mathrm{ss}$ \citep{2005MNRAS.363.1031F}. This is also consistent with the observed form of the column density distribution $f(N_\mathrm{HI}, z)$ at high $N_\mathrm{HI}$ \cite{2007ApJ...656..666O,2009ApJ...696.1543P,2009ApJ...705L.113P,2010ApJ...718..392P,2011ApJ...743...82M,2021MNRAS.501.5811K}. With these considerations in mind, we assume the following:
\begin{enumerate}

    \item For every cell $i$, there exists a turn-over density $\Delta_{t, i}$ beyond which the conditional PDF is of a power-law form, i.e., 
          \be
          P_V(\Delta | \Delta_i) = \mathcal{N}_V(\Delta_i) \, \Delta^{-\beta_V} \quad \text{for} \quad \Delta > \Delta_{t, i},
          \label{eq:P_V_power_law}
          \ee
          where $\mathcal{N}_V(\Delta_i)$ is a normalization factor and $\beta_V$ is the power-law index. The turn-over happens for moderate densities for the unconditional distribution \cite{2005MNRAS.363.1031F}.

    \item The self-shielded threshold density $\Delta_{\mathrm{ss}, i} \gg \Delta_{t, i}$ for all $i$. Thus the power-law form of the conditional distribution is established at $\Delta \sim \Delta_{\mathrm{ss}, i}$.

\end{enumerate}

With these assumptions, it is straightforward to write the clumping factor using \eqn{eq:C_H_i} as\footnote{Hydrodynamical simulations indicate that gas can be depleted from regions with densities $\sim \Delta_{\mathrm{ss},i}$ as a consequence of dynamical processes, which then implies that the commonly adopted power-law scaling of the density PDF is applicable only for densities exceeding this threshold \cite{2020ApJ...898..149D}. In this case $\Delta_{\mathrm{ss},i} \approx \Delta_{t,i}$, and the assumption of a simple power-law dependence for $C_{H,i}$ breaks down. Under these conditions, an accurate description requires extending the model to incorporate the complete density probability distribution function rather than relying on a single-slope approximation.}
\be
C_{H,i} = \f{\mathcal{N}_V(\Delta_i)}{\Delta_i^2 (3 - \beta_V)}~\Delta_{\mathrm{ss}, i}^{3 - \beta_V} \qquad (\beta_V < 3).
\ee
For the normalization, in absence of any other inputs, we assume a power-law dependence on density and redshift
\be
\mathcal{N}_V(\Delta_i) = \mathcal{N}_{V, 0} \, (3 - \beta_V) \, \Delta_i^{2 + \gamma_V} \, \left(\frac{1 + z}{6.5}\right)^{-\alpha_V},
\label{eq:N_V_Delta_i}
\ee
so that the clumping factor becomes
\be
C_{H,i} = \mathcal{N}_{V, 0}~\Delta_i^{\gamma_V}~\Delta_{\mathrm{ss}, i}^{3 - \beta_V} \, \left(\frac{1 + z}{6.5}\right)^{-\alpha_V}.
\label{eq:clumping}
\ee
In our work, $\mathcal{N}_{V, 0}$, $\gamma_V$, $\alpha_V$ and $\beta_V$ are free parameters, to be fixed by comparing the model predictions with the observations.

With these assumptions, the expression for the mean free path, given by \eqn{eq:lambda_ss_i_P_V},  too simplifies to
\be
\f{1}{\lambda_{\mathrm{ss}, i}} = \f{\mathcal{N}_V(\Delta_i)}{f_s \, \lambda_0 \, (\beta_V - 3/2)} \left(\f{T_{\mathrm{HII}, i}}{10^4~\text{K}}\right)^{-1/2}~\Delta_{\mathrm{ss}, i}^{3/2 - \beta_V} \qquad (\beta_V > 3/2),
\ee
where we have assumed that the temperature $T(\Delta)$ is dominated by the temperature around the self-shielded threshold density $\Delta_{\mathrm{ss}, i}$, which in turn is simply the temperature $T_{\mathrm{HII}, i}$ of the ionized regions in the cell.

In fact, we can eliminate the normalization factor $\mathcal{N}_V(\Delta_i)$ from the above expression and write the mean free path in terms of the clumping factor as
\be
\lambda_{\mathrm{ss},i} = f_s~\lambda_0 ~\left(\f{T_{\mathrm{HII}, i}}{10^4~\text{K}}\right)^{1/2} \f{\Delta_{\mathrm{ss}, i}^{3/2}}{C_{H,i}~\Delta_i^2},
\label{eq:lambda_ss}
\ee
where we have absorbed a factor of $(\beta_V - 3/2) / (3 - \beta_V)$ into the unknown normalization factor $f_s$. Note that these relations are valid only for $3/2 < \beta_V < 3$. Since the self-shielded threshold density $\Delta_{\mathrm{ss}, i}$ depends on the photoionization rate, our model naturally captures the interplay between ionizing emissivity, recombinations, and mean free path evolution.

To see this, let us write the self-shielded threshold density in terms of the photoionization rate using \eqn{eq:Delta_ss} and obtain
\be
C_{H,i} = f_s~ \f{\Gamma_{\mathrm{HI}, i}}{\sigma_\mathrm{HI}(\nu_\mathrm{HI})~\lambda_{\mathrm{ss}, i}} \f{1}{\alpha_A(T_{\mathrm{HII}, i})~\chi_\mathrm{He}~n_{H,i}^2~(1+z)^5}.
\label{eq:clumping_Gamma_mfp}
\ee
This relation between the clumping factor, the photoionization rate, and the mean free path can also be obtained by equating the number of recombinations to the number of photoionizations, under the assumption that the IGM is in ionization equilibrium within a region of characteristic size $c / \lambda_{\mathrm{ss},i}$ \cite{2024arXiv240618186D}. Although the expression in \eqn{eq:clumping_Gamma_mfp} was originally derived through a different line of reasoning, it is noteworthy that it yields a form consistent with results from independent studies. 

Extending this argument, the relation can be expressed as
\be
\lambda_{\mathrm{ss}, i} = \tilde{\lambda} \, \f{f_s}{C_{H,i}} \, \f{\Gamma_{\mathrm{HI}, i}}{0.1 \times 10^{-12} \ \mathrm{s}^{-1}} \, \Delta_i^{-2} \left( \frac{1 + z}{7} \right)^{-5},
\ee
where the normalization is $\tilde{\lambda} \approx 19\,\mathrm{cMpc}$ for Case~A recombination, increasing to $\approx 31\,\mathrm{cMpc}$ when adopting Case~B.  

An equivalent scaling relation can also be derived by considering the attenuation of the photon flux through a thin rectangular slab \cite{2013ApJ...763..146E}. In that case, the normalization is found to be $\tilde{\lambda} \approx 73\,\mathrm{cMpc}$ (Case~B), a factor of $\approx 2.4$ larger than our value. This higher normalization agrees well with results from hydrodynamical simulations. The difference between our normalization and the simulation-calibrated value can plausibly be attributed to variations in the assumed spectral indices of both the ionizing sources and the background radiation, combined with a fully consistent treatment of the frequency dependence of the relevant cross-section and radiation fields.  

Crucially, despite being derived without reference to simulation results, our estimate for $\tilde{\lambda}$ is already in close agreement with numerical studies. The remaining discrepancy could be reconciled by adjusting the free parameter $f_s$. A detailed comparison with state-of-the-art hydrodynamical simulations, and a systematic exploration of the parameter space, will be presented in future work.

It is clear from the discussion above that our modeling of sub-grid physics leads to fluctuating clumping factor, mean free path, and hence photoionization rate in the simulation volume.

\subsection{Observables and physical quantities}

Our model enables the computation of a wide range of observables and physical quantities, as outlined below:

\begin{itemize}

    \item The UV luminosity function (UVLF) at various redshifts can be computed using equations \eqref{eq:Phi_UV_cell} and \eqref{eq:Phi_UV_global}. This can be compared with observations from HST \cite{2021AJ....162...47B} and JWST \cite{2023MNRAS.518.6011D,2023ApJS..265....5H,2023MNRAS.523.1009B,2024MNRAS.527.5004M,2024MNRAS.533.3222D}.

    \item The reionization history, characterized by the evolution of the globally averaged ionization fraction $Q_\mathrm{HII} \equiv \langle x_{\mathrm{HII}, i} \Delta_i \rangle$, can be, in principle, compared with constraints from various high-redshift probes \cite{2018MNRAS.479.1055B,2022MNRAS.514...55B,2018ApJ...864..142D,2022MNRAS.512.5390G,2024ApJ...969..162D,2024ApJ...971..124U}.  However, it is important to note that several of these constraints are model-dependent.

    \item The CMB optical depth $\tau_e \equiv \tau_e(z_\mathrm{LSS})$ to the last scattering surface $z_\mathrm{LSS}$ is obtained from
          \be
          \tau_e(z) = \sigma_T \bar{n}_H c \int_0^z \frac{\de z'}{H(z')} (1 + z')^2 \chi_\mathrm{He}(z') Q_\mathrm{HII}(z'),
          \ee
          where $\sigma_T$ is the Thomson cross-section. The value of $\tau_e(z_\mathrm{LSS})$ can be compared with measurements from CMB anisotropies, e.g., Planck \citep{2020A&A...641A...6P}.

    \item The globally-averaged photoionization rate, defined as $\langle \Gamma_{\mathrm{HI}, i} \rangle$, can be compared with constraints obtained by comparing hydrodyanmical simulations with Ly$\alpha$ forest measurements \citep{2011MNRAS.412.1926W,2023MNRAS.525.4093G}. The observable that is more relevant while comparing with the observations is the average within ionized regions, computed as $\Gamma_\mathrm{HI} = \langle \Gamma_{\mathrm{HI}, i} \rangle / \langle x_{\mathrm{HII}, i} \rangle$. Since most observations occur at $z \lesssim 6$, after reionization, the distinction between whether one should use all the regions or only the ionized regions do not make a large difference in results.

    \item Using the thermal history, we calculate the temperature-density relation for the low-density IGM, which is approximated as a power law:
          \be
          T = T_0 \, \Delta^{\gamma - 1},
          \ee
          where $T_0$ is the temperature at the mean density and $\gamma$ is the slope \citep{1997MNRAS.292...27H}. As our grid cells are relatively large, sub-grid modeling techniques, detailed in \citet{2022MNRAS.511.2239M}, are employed to compute $T_0$ and $\gamma$, enabling comparison with Ly$\alpha$ absorption spectra.

    \item The globally averaged mean free path ($\lambda_\mathrm{mfp}$) is defined as:
          \be
          \lambda_\mathrm{mfp} = \frac{\Delta x}{- \ln \left \langle \e^{-\Delta \tau_i} \right \rangle} = \frac{\Delta x}{- \ln \left \langle \e^{-\Delta x/\lambda_{\mathrm{mfp}, i}} \right \rangle}.
          \ee
          This is statistically equivalent to computing the flux decrement as a function of distance and fitting it with an exponential profile \citep{2023ApJ...955..115Z}. However, observed sightlines are toward luminous quasars which reside in high-density peaks of the density field. This may introduce bias in the computation of $\lambda_\mathrm{mfp}$ \citep{2024MNRAS.533..676S}. Since it is not straightforward to model the abundance of quasars without accounting for their properties, we ignore this aspect in this work.

     \item The ionizing emissivity ($\dot{n}_\mathrm{ion}$), defined as $\dot{n}_\mathrm{ion} = \langle \dot{n}_{\mathrm{ion}, i} \rangle$, while not directly observable, is useful for comparing source models with other theoretical predictions. In practice, $\dot{n}_{\mathrm{ion}}$, can however be \textit{indirectly} inferred from measurements of the photoionization rate ($\Gamma_{\mathrm{HI}}$) and the mean free path of ionizing photons  ($\lambda_{\mathrm{mfp}}$) obtained by combining Lyman-$\alpha$ forest observations with detailed hydrodynamical simulations \cite{2007MNRAS.382..325B, 2013MNRAS.436.1023B,  2021MNRAS.508.1853B, 2023MNRAS.525.4093G}.
 
    \item The globally averaged clumping factor ($\mathcal{C}_\mathrm{HII}$) is expressed as
          \be
          \mathcal{C}_\mathrm{HII} = \frac{\langle C_{H, i} \, \Delta_i^2 \, x_{\mathrm{HII}, i} \, (T_{\mathrm{HII}, i} / 10^4 \, \mathrm{K})^{-0.7} \rangle}{\langle x_{\mathrm{HII}, i} \, \Delta_i \rangle}.
          \label{eq:C_HII}
          \ee
          This plays a crucial role in determining the reionization history via:
          \be
          \frac{\de Q_\mathrm{HII}}{\de t} = \frac{\dot{n}_\mathrm{ion}}{\bar{n}_H} - \chi_\mathrm{He} \, \mathcal{C}_\mathrm{HII} \, \bar{n}_H^2 \, Q_\mathrm{HII} \, \alpha_A(T = 10^4 \, \mathrm{K}) \, (1 + z)^3.
          \ee
          Due to the explicit photon conservation in our model, the clumping factor obtained in \eqn{eq:C_HII} by averaging over all cell ensures the equation above is satisfied identically.

\end{itemize}

\subsection{Free parameters}

The model has several free parameters, which can be broadly classified into two categories: those related to the sources and those related to the IGM and sub-grid physics. The \emph{eight} parameters related to the sources are:

\begin{itemize}
    \item $\ell_{\star,0}$, $\ell_{\star,\mathrm{jump}}$, $z_\mathrm{trans}$, $\Delta z$, $\beta_{\star,0}$, $\beta_{\star,\mathrm{jump}}$: These \emph{six} parameters, defined in \eqns{eq:epsilon_star}{eq:epsilon_star_tanh}, determine the quantity $\varepsilon_{\star}(M_h, z)$, which represents a combination of the star-forming efficiency and the specific luminosity of galaxies. These six parameters are sufficient to model the UVLF.

    \item $\varepsilon_{\mathrm{esc},10}$, $\beta_\mathrm{esc}$: These \emph{two} parameters determine the escaping ionizing efficiency, $\varepsilon_{\mathrm{esc}}(M_h)$, as described in \eqn{eq:epsilon_esc}. These parameters are assumed to be independent of redshift.
\end{itemize}

The \emph{six} parameters related to the IGM, including sub-grid physics, are:
\begin{itemize}
    \item $T_\mathrm{re}$: The reionization temperature, as defined in \eqn{eq:epsilon_PH}.

    \item $\mathcal{N}_{V, 0}$, $\gamma_V$, $\alpha_V$: These \emph{three} parameters control the amplitude of the conditional density PDF $P_V(\Delta | \Delta_i)$, as given in \eqn{eq:N_V_Delta_i}.

    \item $\beta_V$: The power-law index of the high-density tail of the conditional density PDF, described in \eqn{eq:P_V_power_law}.

    \item $f_s$: The normalization of the mean free path, as defined in \eqn{eq:lambda_ss_i_P_V}.
\end{itemize}

In total, the model comprises \emph{fourteen} parameters. While varying these parameters to find the best fit to observational data, we use the logarithm of $\varepsilon_{\mathrm{esc},10}$, $T_\mathrm{re}$, $\mathcal{N}_{V, 0}$, and $f_s$, as these parameters are positive definite and can vary over several orders of magnitude.

Having established our theoretical framework and its sub-grid model components, the following section examines the model's predictive capability by comparing its outcomes directly with observational data and exploring the sensitivity of our predictions to various model parameters.

\section{Model Comparison with the Data}
\label{sec:model_comparision_and_dependence}

The primary goal of this study is not only to develop a robust theoretical framework but also to validate its applicability against observational data. This section systematically compares predictions from our fiducial model with various observational datasets. Rather than extensively exploring the parameter space -- which is reserved for future work -- we concentrate on elucidating the model’s physical properties and understanding the sensitivity of observables to the parameters, especially those associated with sub-grid physics. Thus, our primary aim is to assess the implications of our modeling choices rather than strictly constraining the reionization history.

\subsection{Data sets and the fiducial model}

The fiducial model is chosen to provide a good fit to the data. The observations used to select the fiducial model are as follows:

\begin{itemize}
    \item UV luminosity function (UVLF): We use UVLF measurements from six redshift bins spanning $5 \leq z \leq 13.2$, based on data from Hubble Space Telescope (HST) and James Webb Space Telescope (JWST) surveys. We use one of the most comprehensive compilations of the UVLF from HST at $z < 9$ \cite{2021AJ....162...47B}. JWST data include analyses from Early Release Observations (ERO) and Early Release Science (ERS) programs, such as CEERS and GLASS \cite{2023MNRAS.518.6011D,2023ApJS..265....5H,2023MNRAS.523.1009B,2024MNRAS.527.5004M} and also a combination of several major Cycle-1 JWST imaging programmes \cite{2024MNRAS.533.3222D,2024MNRAS.527.5004M}. To minimize uncertainties from physical processes not accounted for in our model, we consider only relatively faint galaxies ($M_\mathrm{UV} \geq -21$), excluding brighter galaxies that are likely to be affected by active galactic nuclei (AGN) feedback or severe dust attenuation \cite{2023MNRAS.526.2196M}.

    \item CMB optical depth: We use the Planck measurement \cite{2020A&A...641A...6P}, which gives $\tau_e = 0.054 \pm 0.007$.

    \item Photoionization rate: Measurements of the photoionization rate $\Gamma_\mathrm{HI}$ at $z \sim 5-6$ are obtained from comparing hydrodynamical simulations with the Ly$\alpha$ forest spectra \cite{2011MNRAS.412.1926W}. Note that there exist more recent measurements that use more sophisticated simulations of Ly$\alpha$ absorption and reionization to determine $\Gamma_\mathrm{HI}$ \cite{2023MNRAS.525.4093G}. However, these measurements are tied to a reionization history that is obtained by fitting other data sets, and using those measurements may cause our inference to be tied towards those constraints. We want to check in the future the reionization constraints from our model independent of other models, hence we choose a measurement that is relatively independent of the reionization history.

    \item Temperature at mean density and slope of the temperature-density relation: We use measurements of $T_0$ and $\gamma$ from Ly$\alpha$ absorption spectra in the range $5.4 \leq z \leq 5.8$ \cite{2020MNRAS.494.5091G}.

    \item Mean free path of ionizing photons: We use measurements of the ionizing mean free path from Ly$\alpha$ absorption spectra at $5.08 \leq z \leq 5.93$ \cite{2023ApJ...955..115Z}.
\end{itemize}

\begin{figure}
    \centering
    \includegraphics[width=0.97\textwidth]{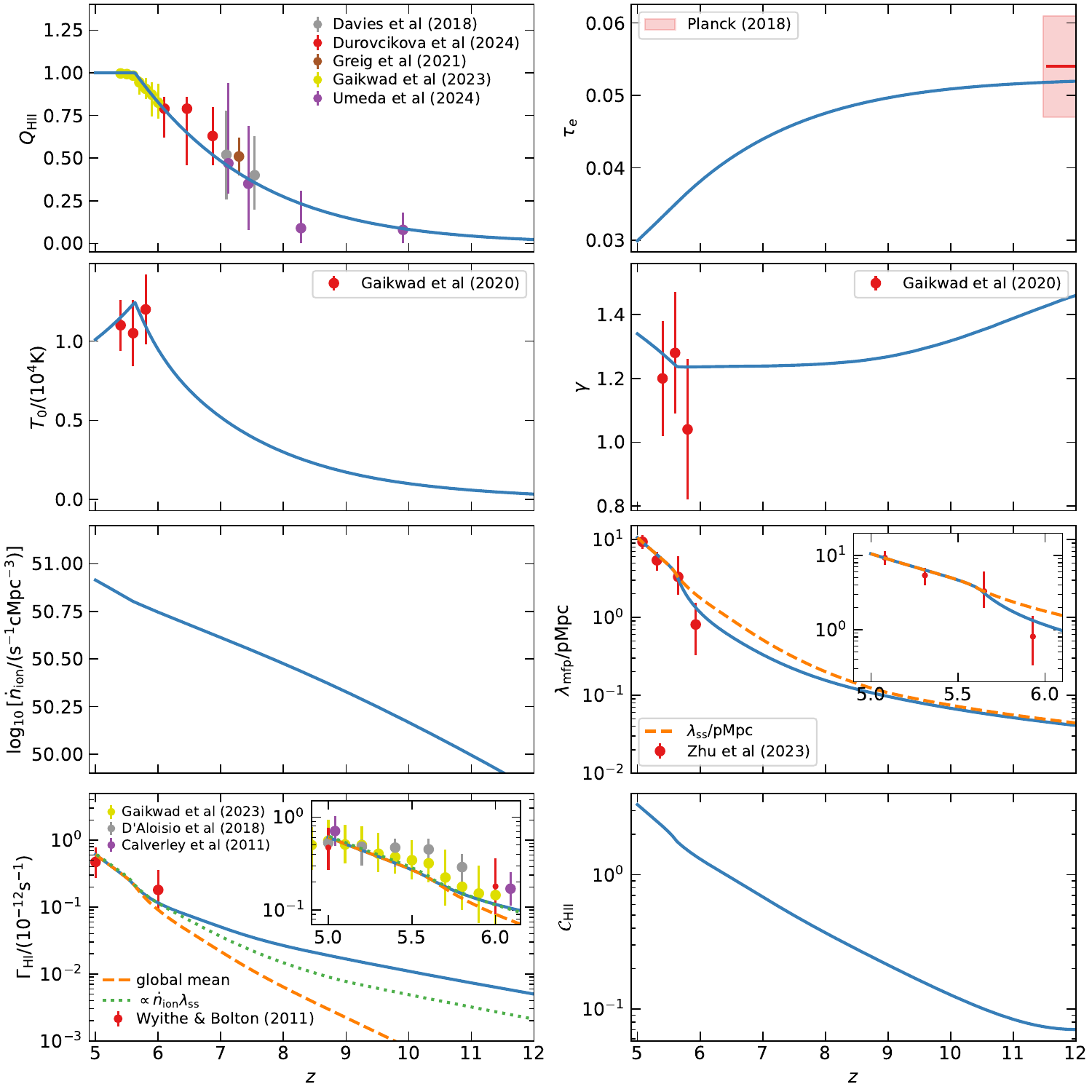}    
    \caption{
        Evolution of globally averaged physical quantities in the fiducial model (blue solid curves), compared with relevant observational constraints. \\
        \textbf{Top:}
        \textit{Left:} Mass-averaged ionized fraction $Q_\mathrm{HII}$. Reionization completes at $z \approx 5.6$ in the fiducial model. Observational constraints (not used in model selection) are shown from Ly$\alpha$ opacity measurements \cite{2023MNRAS.525.4093G}, damping wing analyses of high-$z$ quasars \cite{2018ApJ...864..142D,2022MNRAS.512.5390G,2024ApJ...969..162D}, and JWST observations of UV-bright galaxies \cite{2024ApJ...971..124U}.\\
        \textit{Right:} CMB optical depth $\tau_e$ integrated up to redshift $z$. The red shaded band shows the $1\sigma$ Planck uncertainty \cite{2020A&A...641A...6P}, with the mean marked by the red line.\\
        \textbf{Second row:}
        \textit{Left:} Temperature at mean density $T_0$, compared with Ly$\alpha$ forest measurements \cite{2020MNRAS.494.5091G}.\\
        \textit{Right:} Slope $\gamma$ of the temperature-density relation, also compared with Ly$\alpha$ constraints \cite{2020MNRAS.494.5091G}.\\
        \textbf{Third row:}
        \textit{Left:} Ionizing emissivity $\dot{n}_\mathrm{ion}$.\\
        \textit{Right:} Mean free path $\lambda_\mathrm{mfp}$ compared with Ly$\alpha$ absorption spectra measurements \cite{2023ApJ...955..115Z}. Dashed line: mean free path in ionized regions ($\lambda_\mathrm{ss}$). Inset zooms in on the redshift range with data. The divergence between $\lambda_\mathrm{mfp}$ and $\lambda_\mathrm{ss}$ at $z \gtrsim 5.6$ reflects the presence of remaining neutral regions.\\
        \textbf{Fourth row:}
        \textit{Left:} Photoionization rate $\Gamma_\mathrm{HI}$ in ionized regions. Red: Ly$\alpha$ forest data \cite{2011MNRAS.412.1926W}. Dashed: global $\Gamma_\mathrm{HI}$ (including neutral regions). Dotted: estimated from emissivity and $\lambda_\mathrm{ss}$. Inset: zoom on data range, including other estimates \cite{2011MNRAS.412.2543C,2018MNRAS.473..560D,2023MNRAS.525.4093G} (not used in fit).\\
        \textit{Right:} Global clumping factor $\mathcal{C}_\mathrm{HII}$, averaged over all grid cells. If Case B recombination is used instead of the default Case A, the clumping factor increases by a factor of $\approx 1.6$, while all other quantities remain unchanged; see \appndx{app:case_B} for details.
    }
    \label{fig:comparison_data_fiducial}
\end{figure}

Additionally, we test the agreement of the fiducial model with Ly$\alpha$ opacity fluctuations \cite{2022MNRAS.514...55B} at $5 \lesssim z \lesssim 6$ in \secn{sec:lya_opacity} and with the constraints on the globally averaged ionization fraction in \fig{fig:comparison_data_fiducial}, however, these data are not used to select the fiducial model.

The fiducial model is identified through a $\chi^2$-based optimization process as outlined below:

\begin{enumerate}

    \item We begin with a low-resolution run ($\Delta x = 16 \hcMpc$, corresponding to $16^3$ grid cells) and perform a $\chi^2$ minimization. To balance contributions from different datasets, the UVLF $\chi^2$ is down-weighted by a factor of $0.1$ which is close to the ratio of points from other data sets to those from the UVLF, ensuring it does not dominate the fit due to the large number of data points.

    \item The best-fit parameters from the low-resolution run are used as initial guesses for a high-resolution run ($\Delta x = 4 \hcMpc$, corresponding to $64^3$ grid cells). We optimize only parameters related to the conditional density distribution ($\mathcal{N}_{V, 0}$, $\gamma_V$, $\alpha_V$ and $\beta_V$) using a Fisher score. The other parameters are kept fixed at their best-fit values from the low-resolution run. The Fisher score-based optimization typically converges within four iterations, yielding the best-fit parameters for the fiducial model.

\end{enumerate}

It is possible that this method may not identify the global minimum of the $\chi^2$ corresponding to the high-resolution simulation as we do not vary all the parameters simultaneously. However, our aim is to find a model that is a good fit to the data. We will explore the parameter space in a future work.

\begin{figure}
    \centering
    \includegraphics[width=0.99\textwidth]{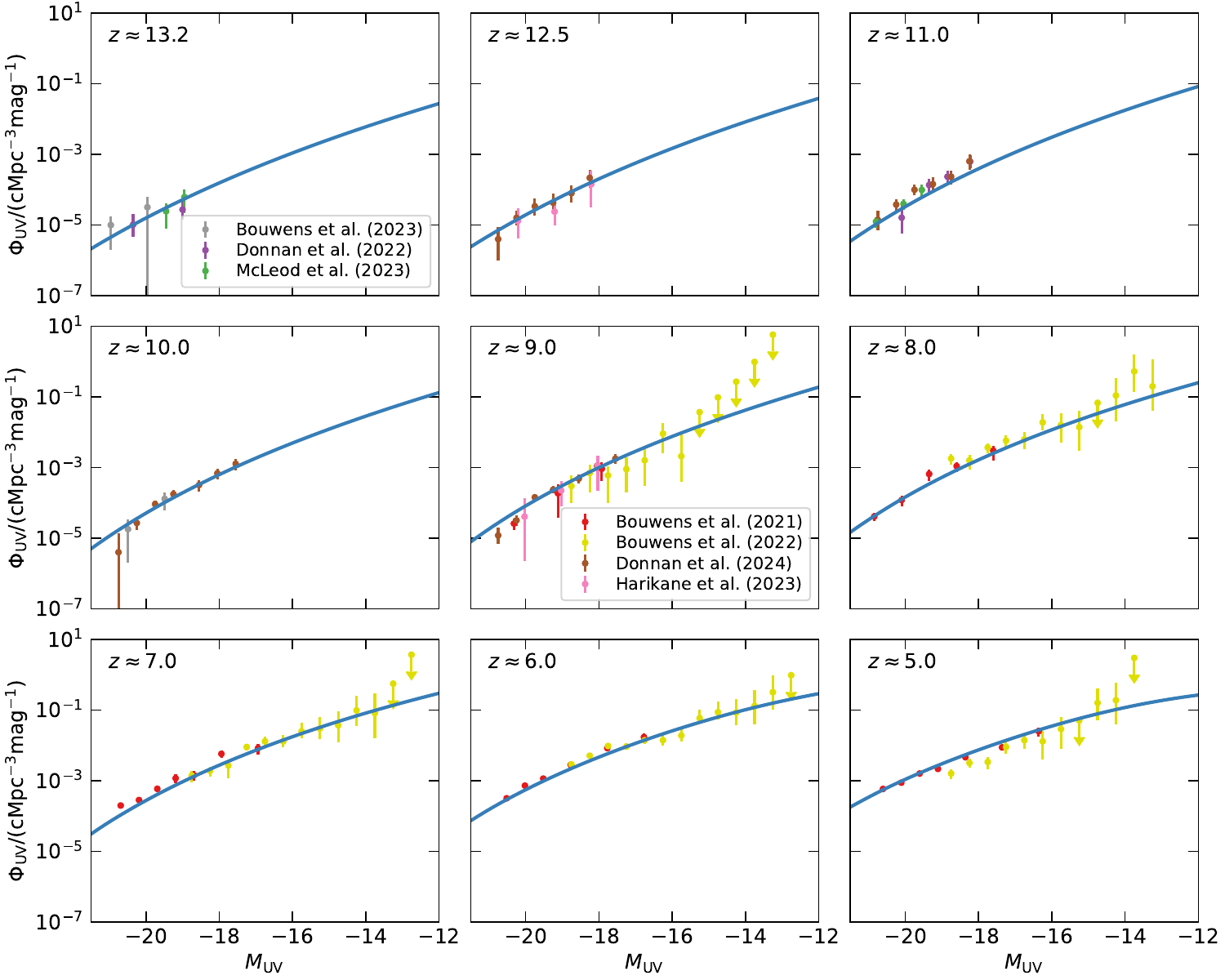}
    \caption{Comparison of the UV luminosity function at various redshifts between the predictions of the fiducial model and observational data represented by points with error-bars. The UVLF data are compiled from different studies that use surveys using
    HST \cite{2021AJ....162...47B} and JWST \cite{2023MNRAS.518.6011D,2023ApJS..265....5H,2023MNRAS.523.1009B,2024MNRAS.527.5004M,2024MNRAS.533.3222D}. We also show data from lensed HFF fields \cite{2022ApJ...940...55B}, yellow points, although they are not used for selecting the fiducial model parameters. The fiducial model provides a good description of the UVLF data across a wide range of redshifts.}
    \label{fig:UVLF_fiducial}
\end{figure}

\begin{table}
    \centering
    \begin{tabular}{|c|c|c|c|c|c|c|c|c|}
        \hline
        Parameters & $\ell_{\star,0}$ & $\ell_{\star,\mathrm{jump}}$ & $z_\mathrm{trans}$ & $\Delta z$ & $\beta_{\star,0}$ & $\beta_{\star,\mathrm{jump}}$ & $\log_{10}\varepsilon_{\mathrm{esc},10}$ & $\beta_\mathrm{esc}$ \\
        \hline
        Values     & $-0.69$          & $5.06$                       & $16.22$            & $7.23$     & $1.82$            & $3.04$                        & $-0.04$                                  & $-0.18$              \\
        \hline
    \end{tabular}

    \vspace{0.2cm}

    \begin{tabular}{|c|c|c|c|c|c|c|}
        \hline
        Parameters & $\log_{10} (T_\mathrm{re} / \mathrm{K})$ & $\log_{10} \mathcal{N}_{V, 0}$ & $\gamma_V$ & $\alpha_V$ & $\beta_V$ & $\log_{10} f_s$ \\
        \hline
        Values     & $4.30$                                   & $-0.33$                        & $-0.02$    & $1.80$     & $2.52$    & $-0.06$         \\
        \hline
    \end{tabular}

    \caption{Parameter values for the fiducial model. The top set corresponds to source parameters, while the middle set represents IGM and sub-grid parameters. See the main text for the definition of the parameters.}
    \label{tab:fiducial_params}
\end{table}

\Tab{tab:fiducial_params} summarizes the parameter values of the fiducial model. Note that the source parameters differ slightly from our previous works \cite{2024JCAP...07..078C} due to differences in the parameterization of star-forming efficiency and $\chi^2$ weighting. Additionally, as mentioned above, the full parameter space for the high-resolution case has not been fully explored in this work, so there may exist other parameter combinations providing similarly good fits.

\Fig{fig:UVLF_fiducial} compares the UVLF at $5 \lesssim z \lesssim 13.2$ predicted by the fiducial model with observational data from HST \cite{2021AJ....162...47B} and JWST \cite{2023MNRAS.518.6011D,2023ApJS..265....5H,2023MNRAS.523.1009B,2024MNRAS.527.5004M,2024MNRAS.533.3222D}. Although we include in the plot data points from lensed HFF fields \cite{2022ApJ...940...55B}, they are not used in the $\chi^2$ minimization and are shown only for comparison. The fiducial model provides a good match to the UVLF data across redshifts, demonstrating the consistency of the star formation efficiency $\varepsilon_{\star, 10}$ and slope $\beta_\star$ with a tanh increase at high redshifts. These findings align with our previous results based on analytical models \cite{2024JCAP...07..078C,2025arXiv250307590C}, and their implications have been discussed extensively in those papers.

\begin{figure}
    \centering
    \includegraphics[width=0.5\textwidth]{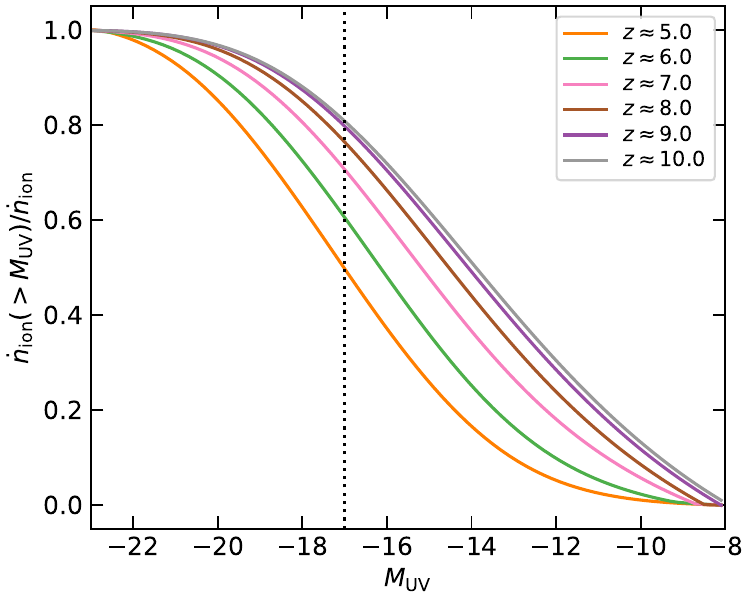}
    \caption{The cumulative distribution of ionizing emissivity produced by galaxies as a function of UV magnitude, shown for different redshifts in the fiducial model. The vertical dashed line indicates $M_\mathrm{UV} = -17$. Most ionizing photons are produced by faint galaxies at all redshifts.}
    \label{fig:ndot_frac_cumulative_fiducial}
\end{figure}

In \fig{fig:ndot_frac_cumulative_fiducial}, we present the cumulative distribution of ionizing emissivity produced by galaxies as a function of UV magnitude for the fiducial model. The results are shown for several redshifts, with vertical dashed line indicating the limiting UV magnitude, $M_\mathrm{UV} = -17$. At all redshifts, most ionizing photons are contributed by galaxies fainter than $M_\mathrm{UV} = -17$, consistent with our earlier works \cite{2024JCAP...07..078C}. This comparison highlights the significant contribution of faint galaxies to the ionizing emissivity, underscoring the importance of including faint-end galaxies in reionization models.

Beyond the source model, we compare the fiducial model with observational data related to the state of the IGM during the reionization epoch. \Fig{fig:comparison_data_fiducial} illustrates the evolution of various physical quantities and compares them with observational constraints wherever possible. The reionization history shows that reionization completes at $z \approx 5.6$ (top row, left panel), consistent with the Planck measurement of the CMB optical depth (top row, right panel). We emphasize that the close agreement between the predicted $Q_{\rm HII}(z)$ from our fiducial model and the observational constraints is achieved despite the fact that these constraints were not used \textit{a priori} in the selection of the fiducial model. The two panels in the second row show that the thermal history produced by the model is consistent with observations at $z \sim 5.5$.

The ionizing emissivity (shown in third row, left panel) increases monotonically with decreasing redshift. The fiducial model does not exhibit strong features from radiative feedback, although a slight change in the slope of the emissivity is visible near the end of reionization due to feedback effects. The monotonic redshift evolution of ionizing emissivity for our fiducial model aligns well with that found in recent fully-coupled simulations, e.g., THESAN \cite{2022MNRAS.512.4909G,2023MNRAS.520.2757Y} and semi-numerical simulations \cite{2024arXiv241200799Q}, but differs from some studies in the literature that report a sharp decline in the emissivity at $z \lesssim 6.5$ \cite{2019MNRAS.485L..24K,2020MNRAS.497..906K,2021MNRAS.507.6108O,2021ApJ...917L..37C,2023MNRAS.525.4093G,2024MNRAS.531.1951C}\footnote{It must however be remembered that the evolution of $\dot{n}_{\mathrm{ion}}$ in most of these studies is not the outcome of a physical model of structure formation, but is instead inferred by tuning the respective simulations to match Ly$\alpha$ forest observations.}. The smooth evolution of $\dot{n}_{\mathrm ion}$ obtained in our case is in fact consistent with the gradual buildup of galaxies implied by observations of galaxy UV LFs at $z < 10$.

The model's prediction for the mean free path $\lambda_\mathrm{mfp}$ is shown in the right panel of the third row and matches measurements from Ly$\alpha$ absorption spectra. At $z \lesssim 5.6$, the mean free path evolves smoothly, almost as a power-law in $1 + z$. However, at $z \gtrsim 5.6$, deviations arise, as is more obvious in the inset panel, due to the presence of neutral regions. These deviations lead to shorter mean free paths, indicative of incomplete reionization at $z \gtrsim 5.6$. For comparison, we also show $\lambda_\mathrm{ss} \equiv -\Delta x / \ln \left \langle \exp \left( - \Delta x / \lambda_{\mathrm{ss}, i} \right) \right \rangle$, the mean free path within ionized regions determined by self-shielded regions. Unlike the global mean free path, $\lambda_\mathrm{ss}$ does not exhibit a dip, as it is unaffected by neutral regions. The value of $\lambda_{\mathrm{mfp}, i}$ as predicted by our fiducial model is broadly in agreement with other theoretical models \cite{2024MNRAS.533..676S,2024arXiv241200799Q}.

The left panel of the bottom row shows the average photoionization rate $\Gamma_\mathrm{HI}$ in ionized regions. The model prediction aligns with Ly$\alpha$ forest measurements, though it evolves slightly more sharply. The global mean photoionization rate is shown as a dashed line and is smaller than the rate in ionized regions at high redshifts. At lower redshifts, where reionization is nearly complete, the two rates converge. Additionally, we show the photoionization rate calculated using a simplified relation involving the ionizing emissivity and the mean free path (taken to be $\lambda_\mathrm{ss}$ for comparison with ionized regions):
\be
\Gamma_\mathrm{HI} \longrightarrow (1 + z)^2 \f{\alpha_s}{\alpha_b + \alpha_\sigma} ~ \sigma_\mathrm{HI}(\nu_\mathrm{HI}) \, \dot{n}_\mathrm{ion} \, \lambda_\mathrm{ss}.
\ee
This approximation holds when $\lambda_\mathrm{ss}$ is uniform and substantially smaller than the horizon size. This approximation is a good match to the full calculation of the photoionization rate at the end stages of reionization, though it does not capture the fluctuations in the mean free path. In the inset, we focus on the redshift interval for which independent constraints on $\Gamma_\mathrm{HI}$ exist, derived from alternative methodologies such as the quasar proximity effect \cite{2011MNRAS.412.2543C} and radiative transfer simulations \cite{2018MNRAS.473..560D,2023MNRAS.525.4093G}. We find that our fiducial model exhibits broad consistency with these external measurements, despite the fact that they were not incorporated into the determination of the fiducial parameters.

Finally, the global clumping factor $\mathcal{C}_\mathrm{HII}$ is shown in the right panel of the bottom row. It increases monotonically with decreasing redshift, ranging from $\sim 1$ at $z = 6$ to $\sim 3$ at $z = 5$. These values are similar in magnitude although somewhat smaller than those from radiative transfer simulations \cite{2025MNRAS.539L..18A}. It is important to note that our clumping factor is directly obtained by averaging over grid cells in the simulation, rather than inferred indirectly from the evolution of the ionized fraction \cite{2025MNRAS.539L..18A,2024arXiv241200799Q}. Photon conservation ensures both methods yield consistent results in our model. Our values are substantially smaller than those inferred from observational estimates of photoionization rate and mean free path \cite{2024arXiv240618186D}, primarily due to differences in recombination rates (case B versus case A) and assumptions about the ionizing source spectrum and background radiation. We have verified that switching from Case A to Case B recombination leaves all key observables unchanged if the normalization $\mathcal{N}_{V,0}$ of the density distribution is rescaled by a factor $\approx 1.6^{\,2 \beta_V / 3 - 1}$, corresponding to a factor of $\approx 1.6$ increase in the clumping factor (see \appndx{app:case_B} for details).

Overall, the fiducial model provides a robust description of a wide range of observables. The next steps involve exploring the implications of the model.

\begin{figure}
    \centering
    \includegraphics[width=0.99\textwidth]{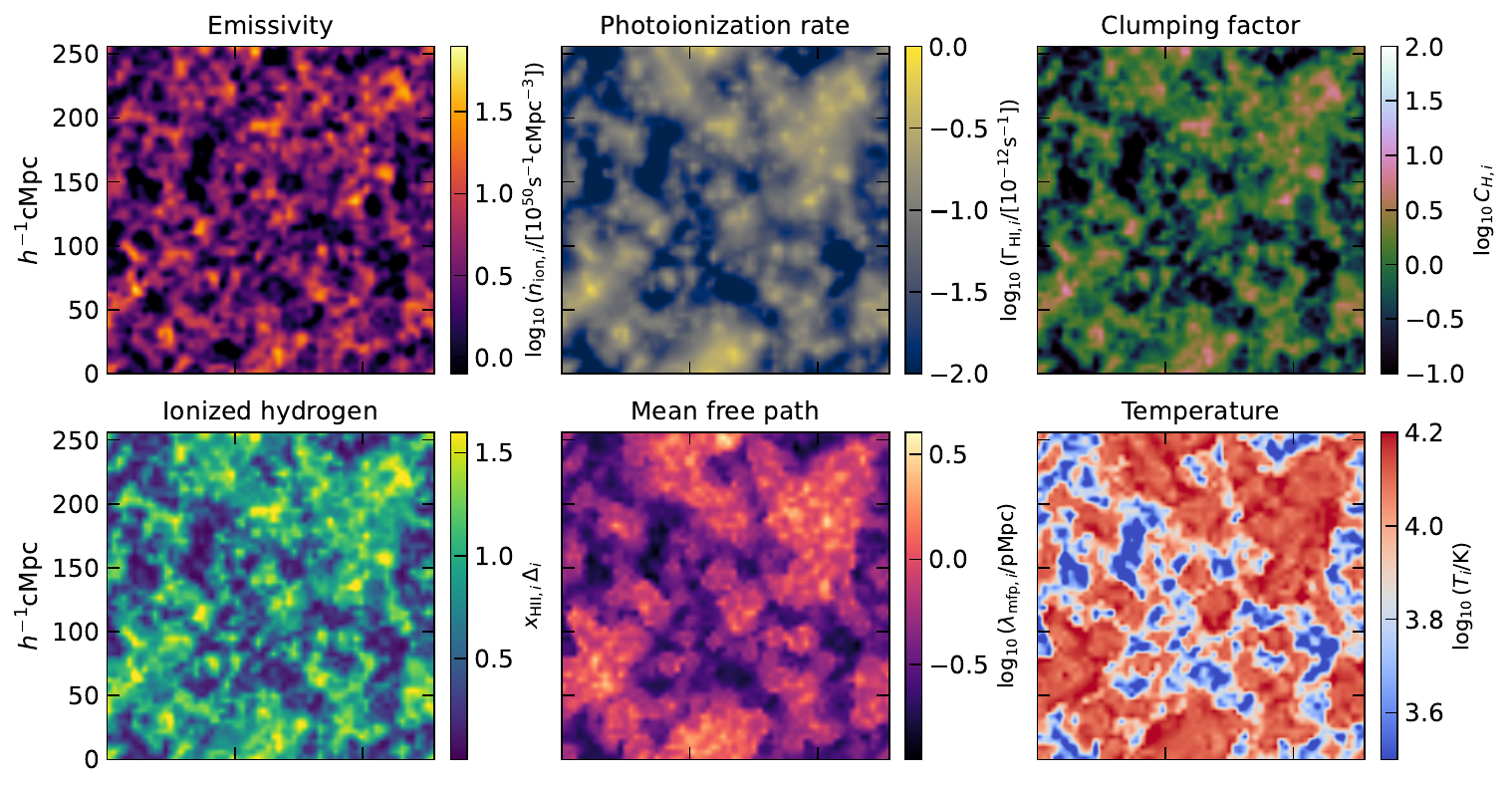}
    \caption{Two-dimensional slices through the simulation volume of the fiducial model at $z = 6$, showing the emissivity, ionized fraction, photoionization rate, mean free path, clumping factor, and temperature. Fluctuations in these quantities are clearly visible, and their inter-correlations are clearly seen.}
    \label{fig:maps_fiducial}
\end{figure}

Our model captures fluctuations not only in the ionization and temperature fields, as shown in our earlier works \cite{2022MNRAS.511.2239M}, but also in the photoionization rate, mean free path, and clumping factor. For visualization, \fig{fig:maps_fiducial} shows a two-dimensional slice through the simulation volume at $z = 6$, plotting various physical quantities. The presence of neutral patches in an otherwise ionized universe leads to large-scale fluctuations that trace the underlying ionization field. Even within ionized regions, quantities like the mean free path, photoionization rate and clumping factor exhibit significant fluctuations due to self-shielded regions.

\begin{figure}
    \centering
    \includegraphics[width=0.99\textwidth]{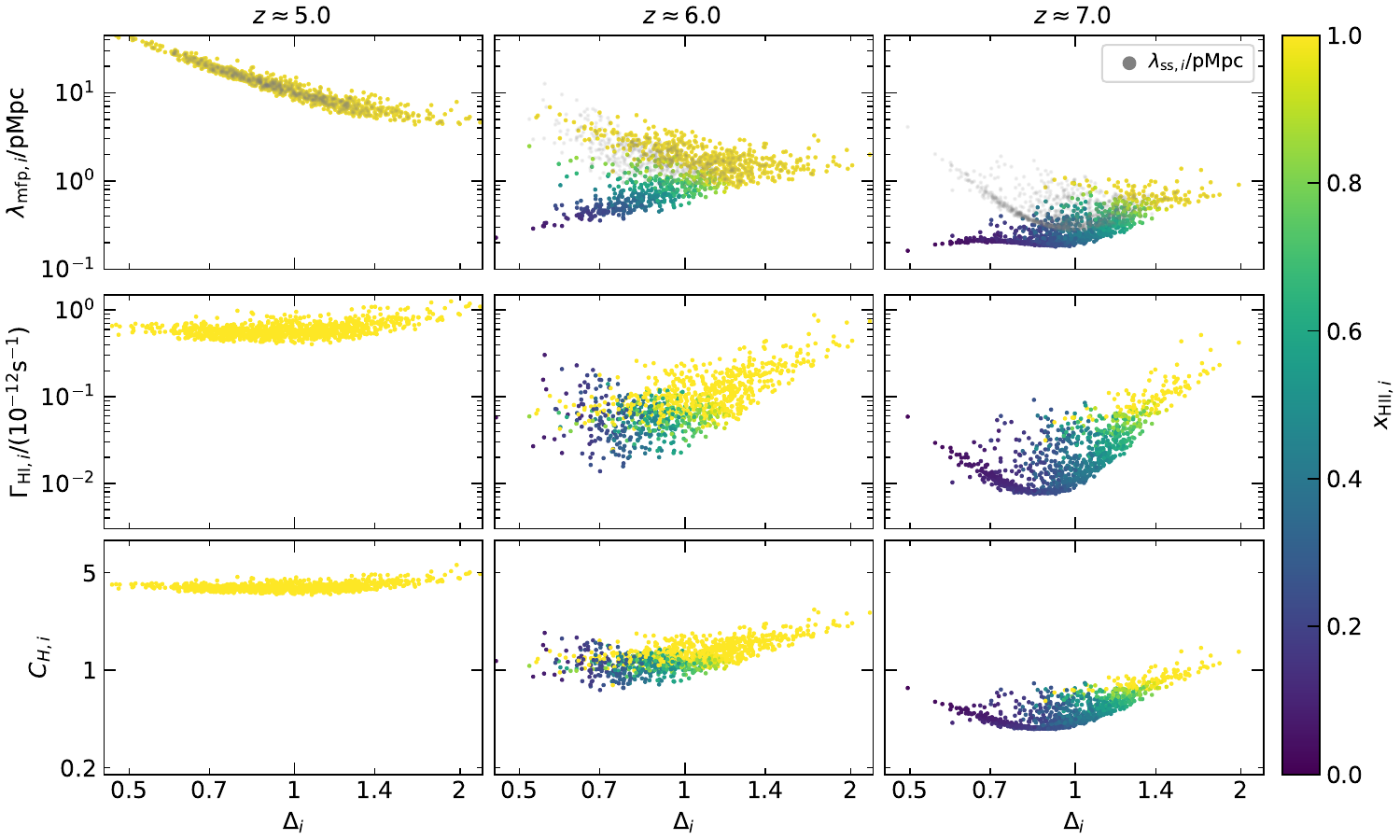}
    \caption{The dependence of physical quantities mean free path $\lambda_{\mathrm{mfp}, i}$ (top row), photoionization rate $\Gamma_{\mathrm{HI}, i}$ (middle row) and clumping factor $C_{H, i}$ (bottom row) on the cell density $\Delta_i$, shown for three redshifts as mentioned in the title of the columns. Each scatter point represents a cell, color-coded by the ionized fraction. For the mean free path,  we also show the self-shielded mean free path $\lambda_{\mathrm{ss}, i}$ by gray points for comparison. It is obvious that $\lambda_{\mathrm{mfp}, i} = \lambda_{\mathrm{ss}, i}$ for highly ionized cells (yellowish points). Although post-reionization $z \sim 5$, the quantities have a clear correlation with the density with susbtantially lower scatter, the pre-reionization redshifts show a more complex relationship due to the presence of neutral islands.}
    \label{fig:density_scaling_fiducial}
\end{figure}

To investigate the relationship between these quantities and the cell density, we present results for three redshifts ($z = 5, 6, 7$) in \fig{fig:density_scaling_fiducial}. The points are color-coded based on the ionized fraction of the cell. The top panel shows the mean free path $\lambda_{\mathrm{mfp}, i}$ as a function of the cell density $\Delta_i$. At $z = 5$, after reionization is complete, the mean free path exhibits an almost one-to-one relationship with the cell density, decreasing as the cell density increases. This is a direct consequence of higher opacity in regions of high-density. At higher redshifts, before reionization completes, low-density regions are not fully ionized, resulting in higher optical depth and an almost constant mean free path for $\Delta_i \lesssim 1$. For comparison, the mean free path $\lambda_{\mathrm{ss}, i}$ due to self-shielded regions is also shown. At high densities, $\lambda_{\mathrm{ss}, i}$ is larger due to higher photoionization rates leading to an increased threshold density for self-shielding. At low densities, $\lambda_{\mathrm{ss}, i}$ is also larger due to reduced opacity.

The middle panel shows the photoionization rate $\Gamma_{\mathrm{HI}, i}$ as a function of density. At $z = 5$, once reionization is complete, the scatter in $\Gamma_{\mathrm{HI}, i}$ is significantly smaller compared to higher redshifts. Before reionization completes, the photoionization rate is higher in high-density cells due to their proximity to sources. The rate decreases with density until it starts to rise again for $\Delta_i \lesssim 1$, where lower opacity leads to higher mean free paths and, consequently, higher photoionization rates.

The clumping factor, shown in the bottom panel, follows a similar trend. At low redshifts, it is nearly independent of density, while at high redshifts, its value is driven by the self-shielded threshold density, which has a larger value in both low- and high-density cells.

This analysis demonstrates that the mean free path, photoionization rate, and clumping factor in ionized regions are intricately linked. An increase in the photoionization rate raises the self-shielded density threshold, which in turn increases the mean free path and subsequently enhances the photoionization rate. This runaway process is mitigated by the corresponding increase in the clumping factor, which amplifies recombinations, thereby increasing opacity and reducing the mean free path. These interrelations are evident in \eqn{eq:clumping_Gamma_mfp}, where we find $\lambda_{\mathrm{ss}, i} \propto \Gamma_{\mathrm{HI}, i} \, C_{H, i}^{-1}$.

\begin{figure}
    \centering
    \includegraphics[width=0.99\textwidth]{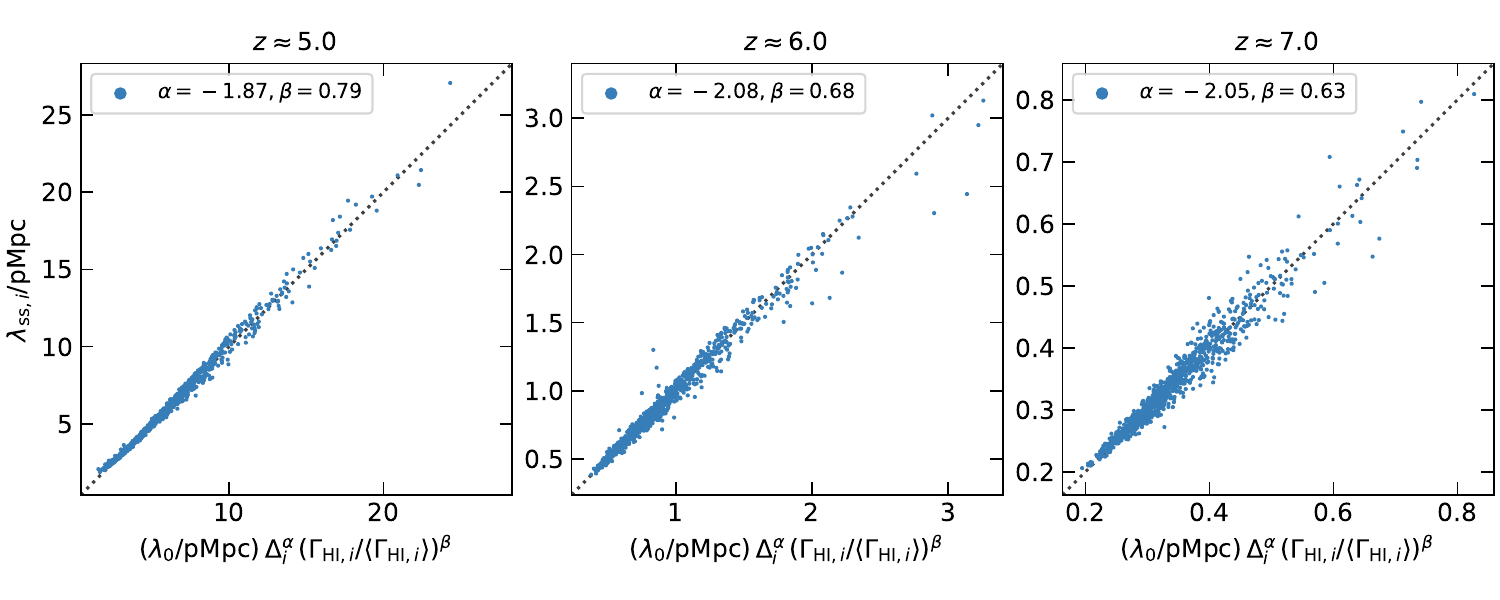}
    \caption{Scaling of the self-shielded mean free path $\lambda_{\mathrm{ss}, i}$ with $\Delta_i^\alpha \, \Gamma_{\mathrm{HI}, i}^\beta$. The scatter points are computed using PCA to determine $\alpha$ and $\beta$. For the fiducial model, $\lambda_{\mathrm{ss}, i} \propto \Delta_i^{-2} \, \Gamma_{\mathrm{HI}, i}^{3/4}$ at $z = 5$ (post-reionization), while at higher redshifts, it scales closer to $\Delta_i^{-2} \, \Gamma_{\mathrm{HI}, i}^{2/3}$.}
    \label{fig:mfp_scaling_fiducial}
\end{figure}

The relationship between the photoionization rate $\Gamma_{\mathrm{HI}, i}$, the mean free path $\lambda_{\mathrm{ss}, i}$, and the density contrast $\Delta_i$ in \emph{ionized cells} is shown in \fig{fig:mfp_scaling_fiducial}. Specifically, we examine whether $\lambda_{\mathrm{ss}, i}$ can be expressed as $\propto \Delta_i^\alpha \, \Gamma_{\mathrm{HI}, i}^\beta$. We use principal component analysis (PCA) to determine the optimal values of $\alpha$ and $\beta$ that minimize the scatter in the $\lambda_\mathrm{ss} - \Delta_i^\alpha \, \Gamma_{\mathrm{HI}, i}^\beta$ plane. At $z = 5$, the mean free path scales as $\lambda_\mathrm{ss} \propto \Delta^{-2} \Gamma_\mathrm{HI}^{3/4}$, while at higher redshifts, the scaling is closer to $\Delta^{-2} \Gamma_\mathrm{HI}^{2/3}$.

These scalings can be understood from the equations derived earlier, namely, \eqnsthree{eq:Delta_ss}{eq:clumping}{eq:lambda_ss}, which give the dependences
\bear
\Gamma_{\mathrm{HI}, i} & \propto \Delta_{\mathrm{ss}, i}^{3/2},
\nline
C_{H, i} & \propto \Delta_i^{\gamma_V} \, \Delta_{\mathrm{ss}, i}^{3 - \beta_V},
\nline
\lambda_{\mathrm{ss}, i} & \propto \frac{\Delta_{\mathrm{ss}, i}^{3/2}}{C_{H, i} \, \Delta_i^2} \propto \Delta_{\mathrm{ss}, i}^{-3/2 + \beta_V} \, \Delta_i^{-2 - \gamma_V},
\ear
where we have neglected the mild temperature dependence of the quantities. Manipulating these equations yields:
\be
\lambda_{\mathrm{ss}, i} \propto \Delta_i^{-2 - \gamma_V} \, \Gamma_{\mathrm{HI}, i}^{-1 + 2 \beta_V / 3},
\ee
which for our fiducial model with $\gamma_V \approx 0$ and $\beta_V \approx 5/2$ becomes
\be
\lambda_{\mathrm{ss}, i} \propto \Delta_i^{-2} \, \Gamma_{\mathrm{HI}, i}^{-3/5}.
\ee

Deviations from this scaling, as well as the scatter in the relation observed in simulations, arise from mild temperature dependencies, which are neglected here. It should also be noted that $\gamma_V$ characterizes the conditional density distribution at the grid scale and is resolution-dependent. Consequently, the density dependence of the mean free path is also resolution-dependent.

In conclusion, our fiducial model not only matches observational data but also provides scaling relations that can aid in constructing simplified models of ionizing background fluctuations. These can also play an important role in comparing with other simulations.

\subsection{Sensitivity to different parameters}

\begin{figure}
    \centering
    \includegraphics[width=0.99\textwidth]{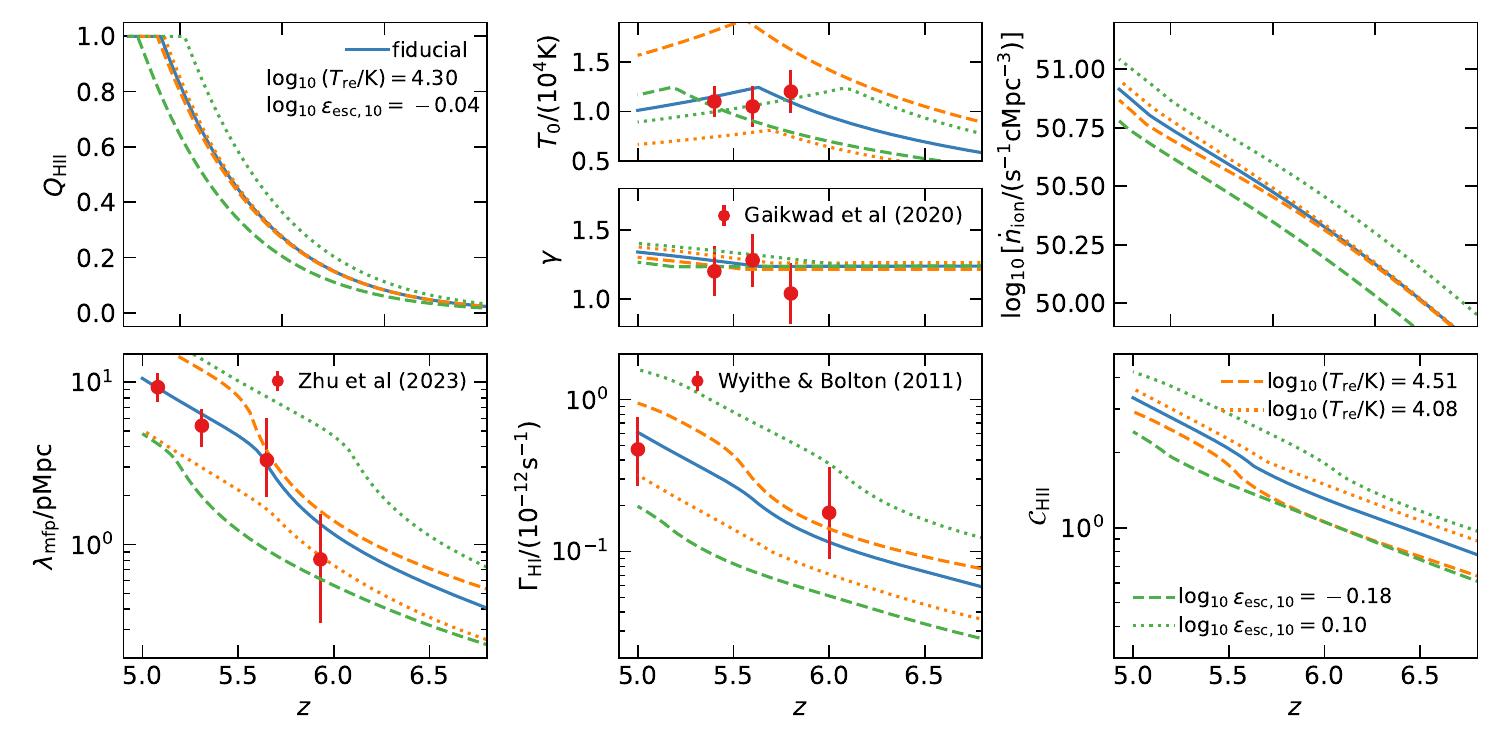}
    \caption{Sensitivity of ionization and thermal histories, and other globally-averaged physical quantities to the reionization temperature $T_\mathrm{re}$ and the amplitude $\varepsilon_{\mathrm{esc}, 10}$ of the escaping ionizing efficiency. Increasing $\varepsilon_{\mathrm{esc}, 10}$ leads to higher ionizing emissivities (top right panel), resulting in earlier reionization (top left panel). Higher emissivities also increases the photoionization rate (bottom middle panel), and that leads to higher mean free path (bottom left panel) and clumping factor (bottom right panel). The main effect of $T_\mathrm{re}$ is to affect the $T_0$ (top middle panel). Higher $T_\mathrm{re}$ also introduces stronger feedback, suppressing ionizing emissivity at lower redshifts (bottom left panel). Furthermore, higher $T_\mathrm{re}$ reduces recombinations, leading to a lower clumping factor (bottom right panel), a longer mean free path (bottom left panel) and a higher photoionization rate (bottom middle panel).}
    \label{fig:parameter_variation_log_T_reion_log_fesc_0}
\end{figure}

We now examine the sensitivity of the predicted quantities to various model parameters, particularly those related to the sub-grid model. These parameters include $\log_{10} (T_\mathrm{re} / \mathrm{K})$, $\log_{10} \mathcal{N}_{V, 0}$, $\alpha_V$, $\gamma_V$, $\beta_V$, and $\log_{10} f_s$. Additionally, we study the impact of varying $\log_{10} \varepsilon_{\mathrm{esc}, 10}$, which determines the amplitude of the escaping ionizing efficiency, to assess its influence on the ionizing emissivity. When varying one parameter, all others are held fixed at their fiducial values. The parameters under consideration are varied by different magnitudes to illustrate more clearly their individual effects on the physical observables. As a result, the following discussion is intended to be qualitative in nature. Due to the non-uniform parameter variations, no quantitative conclusions regarding the sensitivity of observables to these parameters should be drawn from this analysis.

\Fig{fig:parameter_variation_log_T_reion_log_fesc_0} shows the impact of $\log_{10} \varepsilon_{\mathrm{esc}, 10}$ and $\log_{10} (T_\mathrm{re} / \mathrm{K})$ on globally-averaged quantities. It is obvious that increasing $\varepsilon_{\mathrm{esc}, 10}$ raises the ionizing emissivity, leading to an earlier reionization. This earlier timeline also affects the evolution of $T_0$, as earlier ionization results in early onset of photoheating. Interestingly, $\varepsilon_{\mathrm{esc}, 10}$ also influences the clumping factor, mean free path, and photoionization rate. A higher emissivity raises the photoionization rate, increasing the density threshold for self-shielding. This in turn raises the clumping factor, and also extends the mean free path, thus further increasing the photoionization rate. In contrast to simpler reionization models where the clumping factor and emissivity are treated as independent  \cite{2003ApJ...586..693W,2005MNRAS.361..577C, 2011MNRAS.413.1569M,2016MNRAS.460..417S, 2024ApJ...961...50S, 2024JCAP...07..078C}, our model dynamically links the clumping factor to emissivity.

Increasing $T_\mathrm{re}$ introduces stronger feedback, suppressing ionizing emissivity at lower redshifts. Higher $T_\mathrm{re}$ also reduces recombinations, leading to a lower clumping factor, a longer mean free path and hence a higher photoionization rate. While the influence of $T_\mathrm{re}$ on the reionization history is minimal, it significantly impacts $T_0$, consistent with our earlier works \cite{2022MNRAS.511.2239M,2022MNRAS.515..617M}.

\begin{figure}
    \centering
    \includegraphics[width=0.99\textwidth]{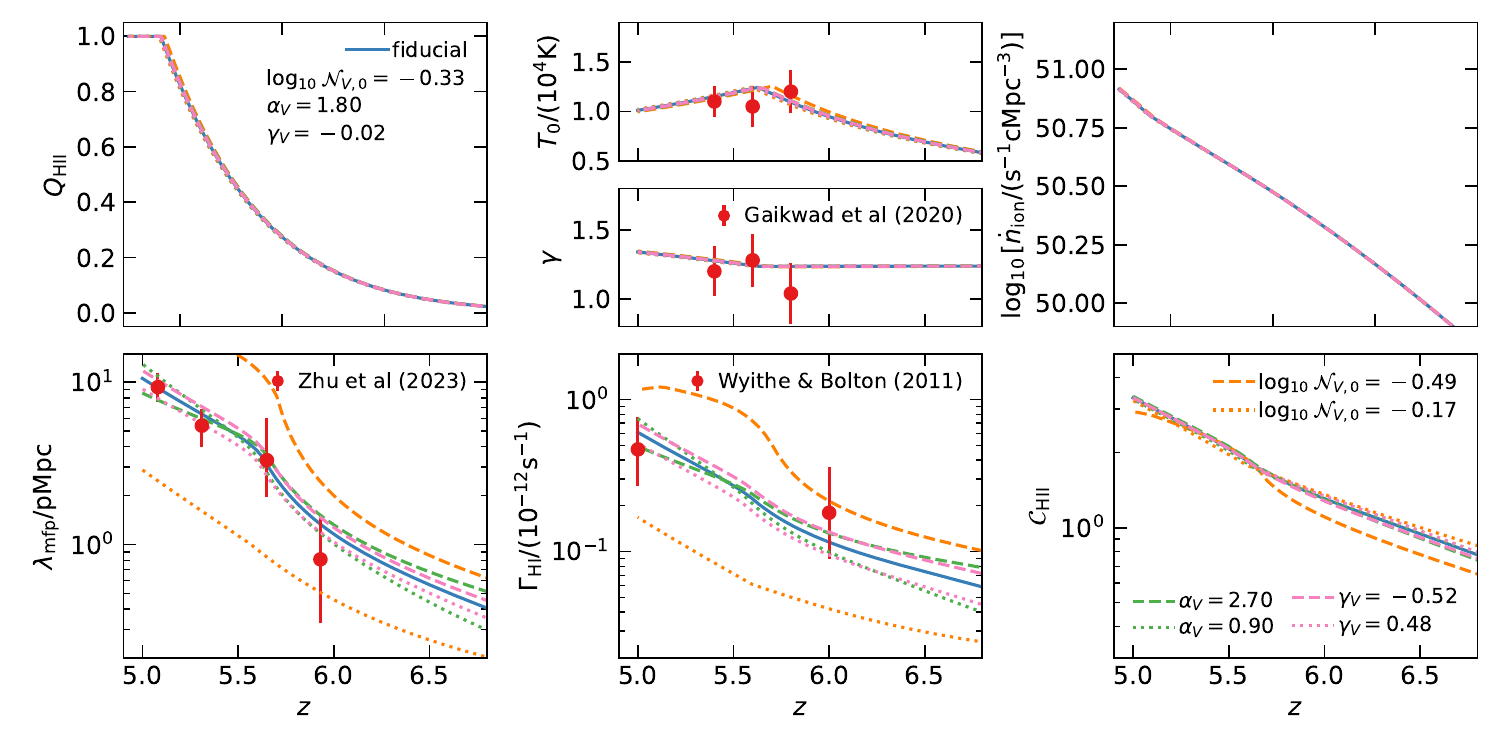}
    \caption{Sensitivity of ionization and thermal histories, and other globally-averaged physical quantities to parameters $\mathcal{N}_{V, 0}$, $\alpha_V$, and $\gamma_V$, which characterize the amplitude $\mathcal{N}_V(\Delta_i)$ of the conditional density distribution. These parameters influence neither the ionization (top left panel) nor the thermal (top middle panel) histories of the universe. Both $\mathcal{N}_{V, 0}$ and $\alpha_V$ affect the abundance of high-density regions, thus affecting the mean free path, photoionization rate and clumping factor (bottom panels). The effect of $\gamma_V$ is negligible on these quantities.}
    \label{fig:parameter_variation_log_norm_clump_0_alpha_clump_gamma_clump}
\end{figure}

In \fig{fig:parameter_variation_log_norm_clump_0_alpha_clump_gamma_clump}, we analyze the effects of varying $\mathcal{N}_{V, 0}$, $\alpha_V$, and $\gamma_V$, which characterize the amplitude $\mathcal{N}_V(\Delta_i)$ of the conditional density distribution. These parameters do not influence the emissivity, reionization history, or thermal parameters $T_0$ and $\gamma$. Increasing $\mathcal{N}_{V, 0}$ raises the abundance of high-density regions, resulting in more high column-density systems and a shorter mean free path. Consequently, the photoionization rate decreases. However, the impact on the clumping factor is minimal, as the increase in the clumping due to increased high-density systems are counterbalanced by a lower self-shielding threshold which reduce clumping.

The parameter $\alpha_V$, which governs the redshift evolution of $\mathcal{N}_V$, decreases $\mathcal{N}_V$ at $z > 5.5$ and increases it at $z < 5.5$ (recall that $z = 5.5$ is our chosen pivot redshift to characterize the evolution of $\mathcal{N}_V$). This behavior explains the dependence of $\lambda_\mathrm{mfp}$ and $\Gamma_\mathrm{HI}$ on $\alpha_V$. An increase in $\gamma_V$ results in an increase in $\mathcal{N}_V$, and consequently its influence on the observables is opposite to that of $\alpha_V$, as shown in the figure.

At this point, it is important to highlight an important aspect of our simulations. It can be seen that for lower $\mathcal{N}_{V, 0}$, the photoionization rate flattens at $z \lesssim 5.5$. This is not because of any physical effect, but because the mean free path approaches the box size (approximately one-third of the box length at this point). This underestimation of the mean free path due to finite simulation volume affects all related quantities. Our analysis highlights the importance of selecting appropriate box sizes, which we explore further in \appndx{app:box_size}. At this point, it is sufficient to highlight that the observed mean free path is significantly smaller than the default box size used in our simulations, making it suitable for parameter estimation.

\begin{figure}
    \centering
    \includegraphics[width=0.99\textwidth]{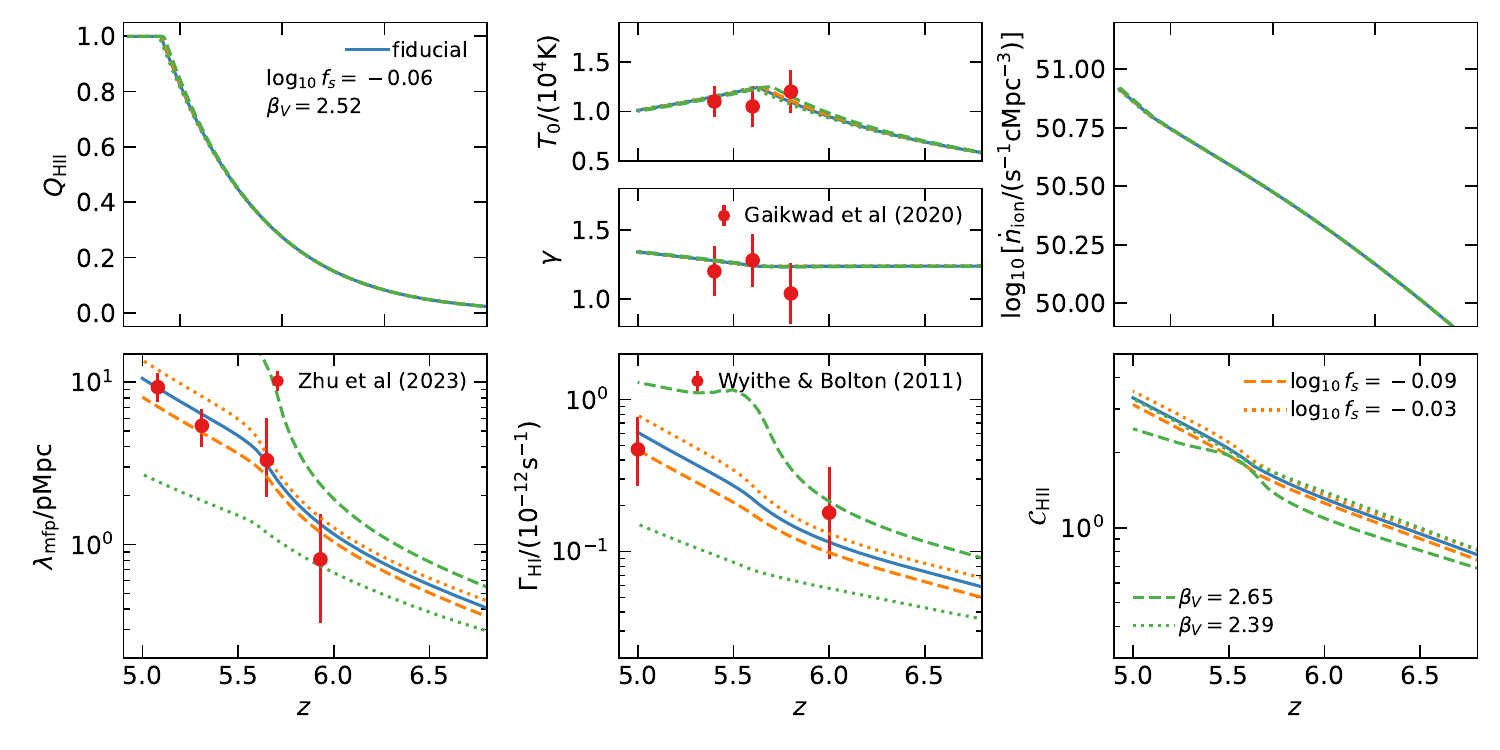}
    \caption{Sensitivity of ionization and thermal histories, and other globally-averaged physical quantities to the slope of the density PDF $\beta_V$ and the normalization of the mean free path $f_s$. None of these parameters affect the ionization (top left panel) and thermal (top middle panel) histories. The parameter $\beta_V$ controls the abundance of high-density regions, affecting the mean free path $\lambda_\mathrm{mfp}$ (bottom left panel) and photoionization rate $\Gamma_\mathrm{HI}$ (bottom middle panel). The normalization $f_s$, directly impacts the same too parameters $\lambda_\mathrm{mfp}$ and $\Gamma_\mathrm{HI}$, as expected.}
    \label{fig:parameter_variation_log_f_s_beta_MHR}
\end{figure}

Finally, \fig{fig:parameter_variation_log_f_s_beta_MHR} explores the effects of varying $\beta_V$ and $\log_{10} f_s$. Similar to the previous figure, these parameters have negligible effects on emissivity, ionization history, and thermal evolution. The slope of the density PDF, $\beta_V$, controls the abundance of high-density regions. A steeper slope reduces the number of high-density systems, increasing the mean free path and photoionization rate. Interestingly, box size effects on $\lambda_\mathrm{mfp}$ and $\Gamma_\mathrm{HI}$ become more pronounced for higher $\beta_V$. The clumping factor decreases marginally with increasing $\beta_V$, as fewer high-density regions form.

The normalization of the mean free path, $f_s$, directly impacts $\lambda_\mathrm{mfp}$ and $\Gamma_\mathrm{HI}$, as expected. Increasing $f_s$ leads to higher values of both quantities, though its effect on the clumping factor is minimal, as it does not directly influence density distributions.

An important aspect that emerges from our analysis is the presence of parameter degeneracies. For instance, while an increase in $\varepsilon_{\mathrm{esc}, 10}$ directly elevates the ionizing emissivity and hence the photoionization rate, similar shifts in the density distribution parameters—such as $\mathcal{N}_{V, 0}$, $\alpha_V$, and $\beta_V$ -- or adjustments to the mean free path normalization $f_s$ can induce comparable changes in both $\Gamma_\mathrm{HI}$ and $\lambda_\mathrm{mfp}$. This overlap in influence implies that distinct combinations of parameters may produce nearly indistinguishable global signatures, complicating the task of uniquely constraining the model. The degeneracies will be studied in detail in future works.

\subsection{The unconditional density distribution}

\begin{figure}
    \centering
    \includegraphics[width=0.6\textwidth]{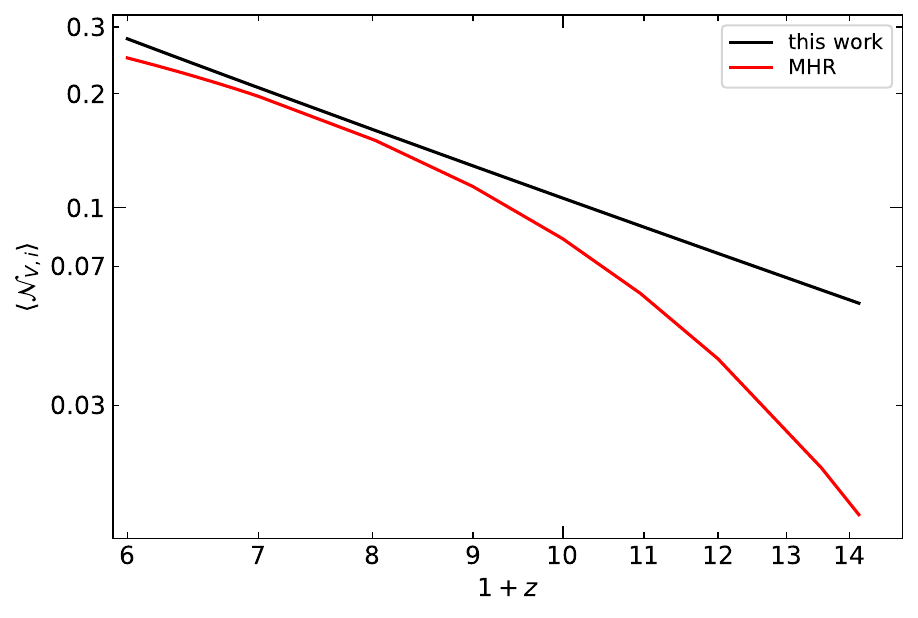}
    \caption{Evolution of the normalization of the unconditional density distribution $P_V(\Delta)$, defined as $\left \langle \mathcal{N}_{V, i} \right \rangle = P_V(\Delta) \, \Delta^{\beta_V}$, as a function of redshift (black curve). The corresponding normalization for the MHR density distribution is also shown in red. At redshifts $z \lesssim 8$, where most of the observational results exist, the two normalizations are in good agreement. At higher redshifts, our simple model evolves differently from the MHR distribution.}
    \label{fig:norm_clump_MHR}
\end{figure}

To understand the parameters that describe the conditional density distribution, which play crucial roles in calculating the clumping factor, mean free path, and photoionization rate, it is instructive to first examine the properties of the corresponding unconditional distribution. We focus on the high-density tail of the distribution ($\Delta \gg 1$), where it can be described by a power-law. Specifically, we consider the quantity
\bear
P_V(\Delta) \, \Delta^{\beta_V} & = \int_0^\infty \de \Delta_i \, P_V(\Delta | \Delta_i) \, P(\Delta_i) \, \Delta^{\beta_V}
\nline
&= \int_0^\infty \de \Delta_i ~ \mathcal{N}_V(\Delta_i) \, P(\Delta_i)
\nline
& \equiv \left \langle \mathcal{N}_{V}(\Delta_i) \right \rangle,
\ear
which represents the globally-averaged normalization factor of the high-density tail. Note that, as this is for the power-law tail of the distribution, this quantity is independent of $\Delta$ and depends only on redshift. \Fig{fig:norm_clump_MHR} shows this quantity for the fiducial model. It follows a power-law dependence on $(1 + z)$, which directly results from our assumption about the redshift evolution, as described in \eqn{eq:N_V_Delta_i}.

We also compute the corresponding normalization for the Miralda-Escudé, Haehnelt, and Rees (MHR) density distribution \cite{2000ApJ...530....1M}, obtained by matching with hydrodyanmaical simulations and also motivated by physical arguments, given by
\be
P_V^\mathrm{MHR}(\Delta) = A_\mathrm{MHR}~\exp\left[-\f{(\Delta^{-2/3} - C_\mathrm{MHR})^2}{2 (2 \delta_\mathrm{MHR} / 3)^2}\right] \, \Delta^{-\beta_\mathrm{MHR}},
\ee
where $A_\mathrm{MHR}$, $C_\mathrm{MHR}$, $\delta_\mathrm{MHR}$, and $\beta_\mathrm{MHR}$ are redshift-dependent parameters. For our analysis, we assume $\beta_\mathrm{MHR} = \beta_V$, which for the fiducial model is $2.52$, nearly identical to the MHR value of $2.5$ at $z = 6$. As outlined in the MHR paper, we adopt $\delta_\mathrm{MHR} = 7.61 / (1 + z)$. The parameters $A_\mathrm{MHR}$ and $C_\mathrm{MHR}$ are determined by normalizing the volume and mass to unity.

Before proceeding, we note that results from more sophisticated hydrodynamical simulations have shown that the MHR model does not fully capture the gas density probability distribution function (PDF), especially at high overdensities \cite{2009MNRAS.394.1812P, 2009MNRAS.398L..26B}. Nonetheless, in the absence of a more accurate analytical alternative, the MHR model remains a useful baseline for comparison.

In the high-density regime ($\Delta \gg C_\mathrm{MHR}$), the MHR distribution simplifies to
\be
P_V^\mathrm{MHR}(\Delta) \approx A_\mathrm{MHR}~\exp\left[- \f{C_\mathrm{MHR}^2}{2 (2 \delta_\mathrm{MHR} / 3)^2}\right] \, \Delta^{-\beta_\mathrm{MHR}}.
\ee
This allows us to compare the normalization $\left \langle \mathcal{N}_{V}(\Delta_i) \right \rangle$ from our fiducial model with the MHR normalization, given by $A_\mathrm{MHR}~\e^{- C_\mathrm{MHR}^2 / [2 (2 \delta_\mathrm{MHR} / 3)^2]}$. The results are shown in \fig{fig:norm_clump_MHR}.

The figure demonstrates that the fiducial model's normalization closely matches the MHR normalization for $5 \lesssim z \lesssim 8$. However, at higher redshifts, the MHR normalization decreases more rapidly compared to our model. This suggests that the power-law redshift evolution assumed in our model may not fully capture the behavior of the MHR distribution at high redshifts. Exploring more physically motivated redshift dependencies is a priority for future work. Nonetheless, it is remarkable that the two models exhibit such close agreement at lower redshifts, especially given that the fiducial model was determined independently of the MHR values.

\subsection{Lyman-\texorpdfstring{$\alpha$}{alpha} opacity fluctuations}
\label{sec:lya_opacity}

\begin{figure}
    \centering
    \includegraphics[width=0.99\textwidth]{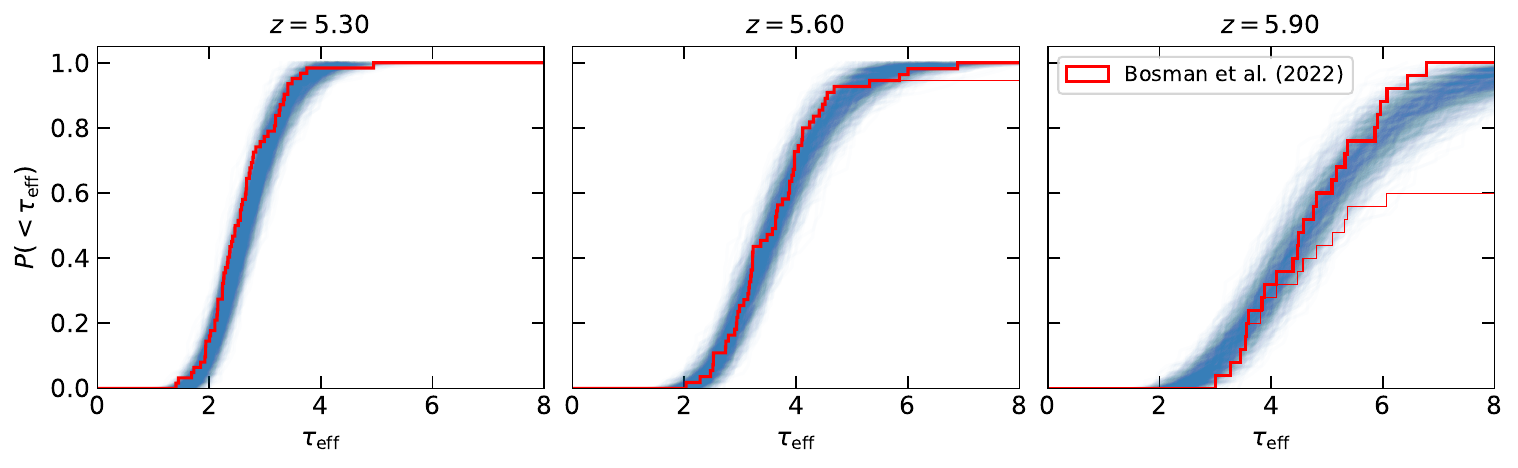}
    \caption{Comparison of the cumulative distribution of the effective Ly$\alpha$ optical depth averaged over a redshift interval $\Delta z = 0.1$ for the fiducial model with observational data at three redshifts that are most relevant to probe the end stages of reionization. The red curves represent the observational data \cite{2022MNRAS.514...55B}, with the upper curves corresponding to lower limits treated as measurements just below the detection sensitivity, and the lower curves corresponding to lower limits treated as $\tau_{\mathrm{eff}} \to \infty$. The blue curves represent the model predictions, with 1000 realizations of the set, each realizations consisting of the same number of sight lines as in the observational data. The agreement between the model and data demonstrates the model's ability to capture the inhomogeneities in the post-reionization IGM and the evolution of the Ly$\alpha$ absorption spectra.}
    \label{fig:tau_eff_lya}
\end{figure}

To calculate the Ly$\alpha$ optical depth, we use the formalism outlined in Choudhury, Paranjape, \& Bosman (2021). Under the fluctuating Gunn-Peterson approximation, the Ly$\alpha$ optical depth in the $i$th grid cell is expressed as
\be
\tau_{\mathrm{GP}, i} = \kappa_{\mathrm{res}} \, \f{\pi \e^2}{m_e c} \, f_{\alpha} \, \lambda_{\alpha} \, \f{(1 + z)^6}{H(z)} \, \chi_\mathrm{He} \, \f{\alpha_B(T_{\mathrm{HII}, i}) \, n_{H, i}^2}{\Gamma_{\mathrm{HI}, i}},
\label{eq:tau_GP}
\ee
where $\kappa_{\mathrm{res}}$ is a normalization factor accounting for small-scale density and velocity fluctuations unresolved by our coarse-resolution simulations \citep{2009ApJ...706..970D,2016MNRAS.460.1328D}, and $f_{\alpha}$ is the Ly$\alpha$ oscillator strength. The remaining symbols have their usual meanings. This equation assumes photoionization equilibrium, making it applicable to fully ionized cells. For $T_{\mathrm{HII}, i}$, we use the ionized region temperature as calculated from \eqn{eq:T_ion}. Note that in this approximation, the computation of the Ly$\alpha$ opacity neglects modifications to the absorption cross-section arising from peculiar velocities of the absorbing gas and from Doppler broadening due to its thermal motions. While the impact of these effects on the mean opacity can, to some extent, be absorbed into the overall normalization parameter $\kappa_{\mathrm{res}}$ in \eqn{eq:tau_GP}, chosen to reproduce the observed distribution of effective Ly$\alpha$ forest optical depths, such a calibration may not fully capture the resulting optical depth distribution \cite{2024arXiv241200799Q}. A systematic investigation of these effects will be undertaken in future work.

For cells containing neutral regions, the optical depth in the ionized portion is computed by replacing $\Gamma_{\mathrm{HI}, i}$ with $\Gamma_{\mathrm{HI}, i} / x_{\mathrm{HII}, i}$, which is equivalent to assuming that the radiation background is concentrated within the ionized regions. Thus, the optical depth in the ionized portion of the cell becomes $x_{\mathrm{HII}, i} \, \tau_{\mathrm{GP}, i}$, where $\tau_{\mathrm{GP}, i}$ is still determined by \eqn{eq:tau_GP}. The optical depth in the neutral part is assumed to be effectively infinite, resulting in zero transmitted flux.

The effective optical depth averaged over $N$ pixels is then given by:
\be
\tau_{\mathrm{eff}} = -\ln \left(\f{1}{N} \sum_{i} x_{\mathrm{HII}, i} \, \e^{-x_{\mathrm{HII}, i} \, \tau_{\mathrm{GP}, i}} \right),
\ee
where $N$ corresponds to the length of the sight lines used in the observational data. This effective optical depth is the key observable we compare with observations.

The observational data used in this work comes from VLT/X-Shooter measurements \citep{2022MNRAS.514...55B}, which provide the Ly$\alpha$ effective optical depth $\tau_{\mathrm{eff}}$ averaged over sightline chunks of varying lengths in the redshift range $5 \lesssim z \lesssim 6$. These data are presented as cumulative distribution functions (CDFs) $P(<\tau_{\mathrm{eff}})$, with two interpretations for lower limits on $\tau_{\mathrm{eff}}$: (i)~lower limits are treated as measurements just below the detection sensitivity, or (ii)~lower limits are assumed to correspond to $\tau_{\mathrm{eff}} \to \infty$. The CDFs are shown in \fig{fig:tau_eff_lya} as red curves, with the upper curves representing the first interpretation and the lower curves the second.

In the same figure, we compare the fiducial model's predictions for the effective Ly$\alpha$ optical depth distribution with observational data at three redshifts. For the model predictions, we use the same number of sight lines as in the observational data and generate 1000 realizations of the set, shown as blue curves. The agreement between the model and data demonstrates the model's ability to capture the inhomogeneities in the post-reionization IGM and the evolution of the Ly$\alpha$ forest.

\section{Summary and Future Outlook}
\label{sec:summary}

In this concluding section, we integrate insights gathered from our theoretical modeling and observational comparisons. We summarize the primary findings of our analysis, emphasizing how our results contribute to the current understanding of reionization. Furthermore, we outline promising directions for future research, particularly emphasizing opportunities opened up by the sub-grid modeling approach introduced in this work.

Understanding the epoch of reionization is crucial for uncovering the astrophysical processes that shaped the early universe. While significant progress has been made in modeling reionization, existing semi-numerical approaches often rely on simplified assumptions about ionizing sources, recombinations, and photon propagation. In this work, we have developed a physically motivated \emph{sub-grid model} within our photon-conserving semi-numerical framework, \texttt{SCRIPT}, to address these limitations. By incorporating spatial fluctuations in key reionization parameters, such as the clumping factor, ionizing mean free path, and photoionization rate, our model captures the complex small-scale physics that governs the ionization state of the IGM. Our model provides a computationally efficient way to capture critical small-scale physics—self-shielded regions, recombinations, and photon sinks—within semi-numerical reionization simulations, thus bringing them closer to the fidelity of computationally expensive radiative transfer methods while maintaining efficiency.

A key advancement of this work is the \emph{explicit coupling of sub-grid physics with the large-scale density field}, allowing for a self-consistent treatment of self-shielded regions and inhomogeneous recombinations. Our model successfully reproduces a wide range of observational constraints, including the UVLF from 
HST~\cite{2021AJ....162...47B} and JWST~\cite{2023MNRAS.518.6011D,2023ApJS..265....5H,2023MNRAS.523.1009B,2024MNRAS.527.5004M,2024MNRAS.533.3222D}, CMB optical depth from Planck~\cite{2020A&A...641A...6P}, and Ly$\alpha$ forest measurements of the IGM temperature~\cite{2020MNRAS.494.5091G}, photoionization rate~\cite{2011MNRAS.412.1926W}, and mean free path~\cite{2023ApJ...955..115Z}. Notably, our model also reproduces the observed Ly$\alpha$ opacity fluctuations~\cite{2022MNRAS.514...55B}, indicating that it accurately captures the patchiness of reionization. Additionally, we have demonstrated that traditionally independent reionization parameters, such as the clumping factor and mean free path, are \emph{strongly correlated}, influencing the timing, morphology, and thermal evolution of reionization. These findings suggest that reionization models lacking such interdependencies may significantly misrepresent the true astrophysical processes at play.

Beyond providing a robust theoretical framework, our results have profound implications for upcoming observational efforts, e.g.,  for interpreting high-redshift galaxy surveys and Ly$\alpha$ forest data. The model's predictive power will be essential for upcoming 21\,cm experiments, which are poised to revolutionize our understanding of reionization. Our ability to self-consistently link ionizing emissivity, mean free path, and recombinations offers a powerful tool for extracting astrophysical parameters from these observations.

Looking ahead, we will extend this framework to incorporate additional physical processes, including inhomogeneous helium reionization and X-ray heating, which are expected to shape the thermal history of the IGM. We also plan to conduct a full Markov Chain Monte Carlo (MCMC) analysis to explore parameter space more systematically and obtain statistically robust constraints on reionization history. Furthermore, with the rapid advancements in computational techniques, we aim to integrate machine learning-based algorithms to accelerate parameter inference and improve predictive capabilities.

In summary, this work provides an advanced semi-numerical framework that bridges the gap between fast but simplistic models and computationally prohibitive radiative transfer simulations. By capturing the essential physics of self-shielded regions and inhomogeneous recombinations, our approach lays a solid foundation for interpreting current and future reionization-era observations. As 21\,cm observations from upcoming telescopes unfold, our framework will provide a vital bridge between simulations and observations, refining our understanding of reionization’s final stages.

\acknowledgments
The authors acknowledge support from the Department of Atomic Energy, Government of India, under project no. 12-R\&D-TFR-5.02-0700. 

\section*{Data Availability}

The data generated during this work will be made available upon reasonable request to the corresponding author.

\appendix

\section{Comparison of Photoionization Rates from SCRIPT and Ray-tracing}
\label{app:ray-tracing} 

\begin{figure}
\centering
\includegraphics[width=0.99\textwidth]{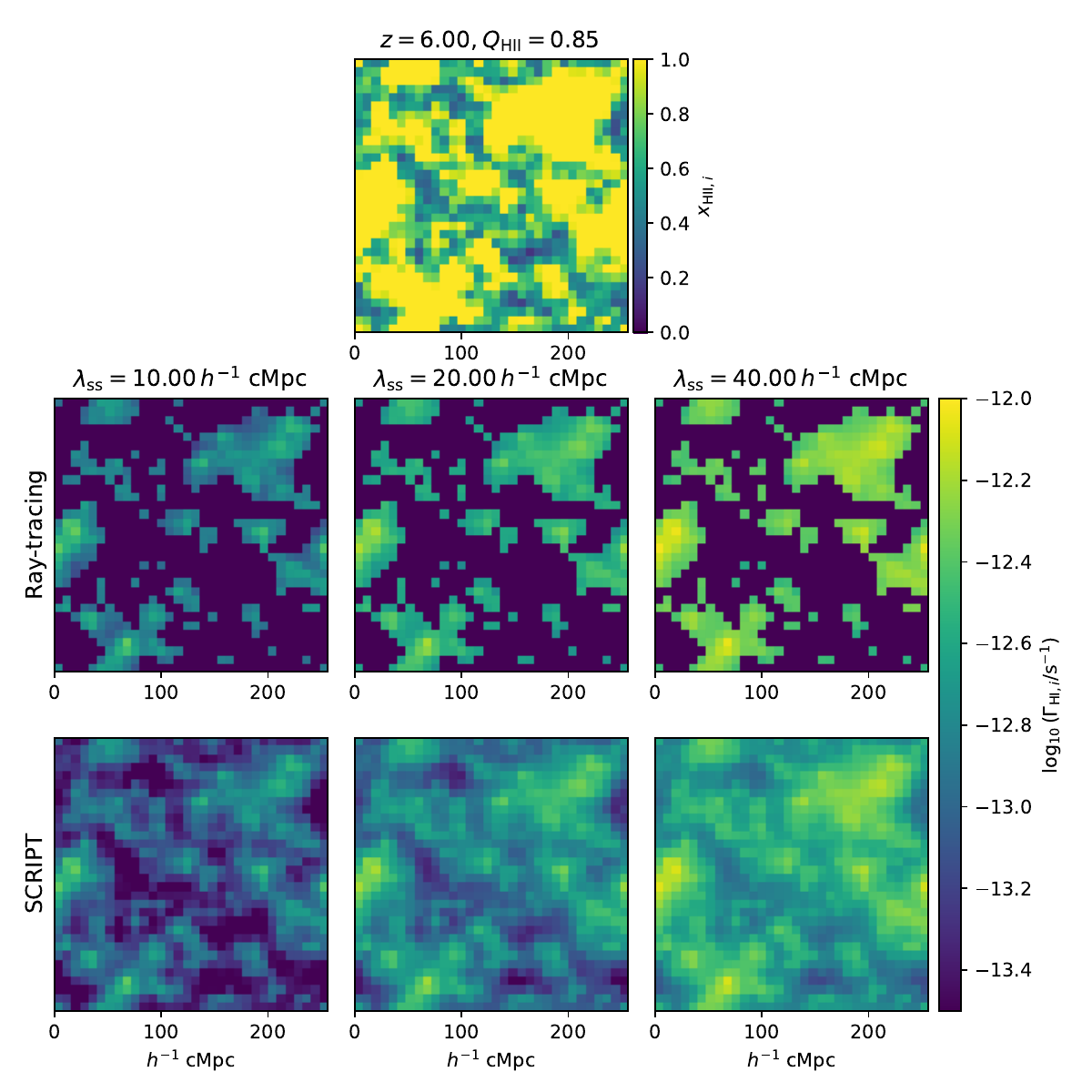}
\caption{Two-dimensional slices through the simulation box illustrating the ionization field and corresponding photoionization rate fields. The top row shows the hydrogen neutral fraction ($x_{\mathrm{HI}, i}$), while the middle and bottom rows show the photoionization rate $\Gamma_{\mathrm{HI}, i}$ computed using the ray-tracing method and SCRIPT, respectively. Results are shown for three different values of the mean free path $\lambda_{\mathrm{ss}}$ in ionized regions. The ray-tracing method captures anisotropic shadowing effects due to neutral absorbers, resulting in more structured $\Gamma_{\mathrm{HI}, i}$ fields, while SCRIPT produces smoother, spherically symmetric profiles due to its isotropic averaging. Differences between the two methods increase with increasing $\lambda_{\mathrm{ss}}$.}
\label{fig:Gamma_HI_cell}
\end{figure}

\begin{figure}
\centering
\includegraphics[width=0.99\textwidth]{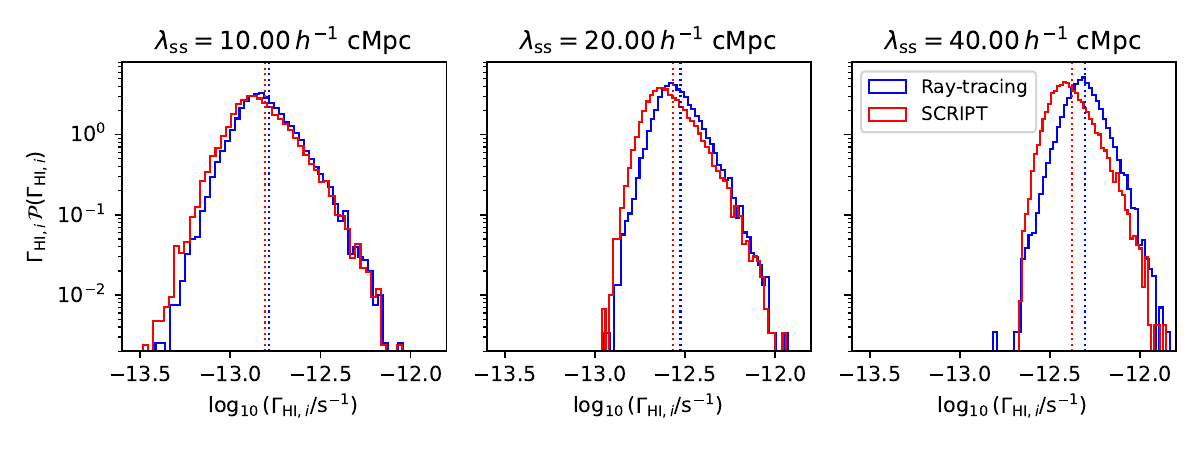}
\caption{Probability distribution functions of the photoionization rate $\Gamma_{\mathrm{HI}, i}$ within ionized regions, comparing results from SCRIPT and the ray-tracing method for three values of $\lambda_{\mathrm{ss}}$. Vertical dashed lines denote the mean $\Gamma_{\mathrm{HI}}$ for each case. The two methods agree closely for small $\lambda_{\mathrm{ss}}$, while discrepancies grow at larger mean free paths, where SCRIPT tends to overestimate $\Gamma_{\mathrm{HI}, i}$ due to its inability to capture directional attenuation from intervening neutral regions.}
\label{fig:Gamma_HI_hist}
\end{figure}

Our method for computing the photoionization rate, $\Gamma_{\mathrm{HI}, i}$, as described in \secn{sec:photoionization}, distributes the ionizing flux over concentric spherical shells. This approach effectively averages the impact of a high-optical-depth absorber located in a specific direction from a source over all directions, thereby enforcing spherical symmetry. To assess the consequences of this spherical averaging on the resulting photoionization rates, we compare our method with a more directional, ray-tracing-based calculation.

In the ray-tracing approach, $\Gamma_{\mathrm{HI}, j}$ is computed directly by summing over contributions from all other cells $i$, using \eqn{eq:Gamma_HI_j}:
\be
\Gamma_{\mathrm{HI}, j} = (1 + z)^2 \f{\alpha_s}{\alpha_b + \alpha_\sigma} ~ \f{\sigma_\mathrm{HI}(\nu_\mathrm{HI})}{4 \pi} \sum_{i \neq j} \dot{N}_{\mathrm{ion}, i}~\f{\e^{-\tau_{i \to j}}}{x_{ij}^2},
\nonumber
\ee
where the optical depth $\tau_{i \to j}$ accounts for absorption along the line of sight through the intergalactic medium (IGM) between cells $i$ and $j$. Owing to this line-of-sight integration, we refer to this approach as ``ray-tracing''. The local contribution of the $i$th cell to $\Gamma_{\mathrm{HI}, i}$ is computed in the same manner as in SCRIPT, following \eqn{eq:Gamma_HI_local}.

This ray-tracing method scales as $\mathcal{O}(N_\mathrm{grid}^2)$, where $N_\mathrm{grid}$ is the total number of grid cells, making it computationally expensive. To mitigate the cost, we perform the ray-tracing comparison on a coarser grid with $N_\mathrm{grid} = 32^3$ (as opposed to our fiducial $64^3$). We also adopt a simplification to further improve computational efficiency: within ionized regions, we assume the mean free path $\lambda_{\mathrm{ss}, i}$ to be spatially uniform. Under this assumption, the optical depth simplifies to
\be
\tau_{i \to j} = \frac{x_{ij}}{\lambda_\mathrm{ss}},
\ee
depending solely on the comoving distance $x_{ij}$ between the two cells. This allows us to avoid explicitly resolving the cell-to-cell intersections along the ray path, thus simplifying the computation.

However, to correctly model shadowing by dense, neutral regions, we introduce a filtering criterion: if any cell $k$ along the line of sight between $i$ and $j$ satisfies $x_{\mathrm{HI}, k} > 0.5$ (i.e., is predominantly neutral), we assume the ionizing photons are completely absorbed and exclude that source-target pair from the sum \cite{2020MNRAS.494.3080N}. While this is a simplification, the inclusion of directionality in the ray-tracing approach allows us to benchmark the limitations of the spherical averaging method.

We first validate both methods in a controlled setup consisting of a single isotropic ionizing source embedded in an otherwise uniform ionized medium, with a single neutral cell placed in a specific direction. As expected, the ray-tracing solution exhibits a "shadow" beyond the neutral cell, whereas our spherical averaging scheme fails to capture this anisotropy, resulting in a spherically symmetric $\Gamma_{\mathrm{HI}}$ field.

The more relevant test, however, involves a realistic cosmological scenario, where the statistical isotropy of source and absorber distributions may mitigate the directional biases. To this end, we adopt the emissivity, density, and ionization fields from our fiducial simulation at a given redshift, and compute the photoionization rate using both the ray-tracing and spherical averaging methods, assuming a fixed value of $\lambda_\mathrm{ss}$.

We present results at redshift $z = 6$, where the mass-averaged ionized fraction is $Q_\mathrm{HII} = 0.85$. We perform the comparison for three representative values of the mean free path, $\lambda_\mathrm{ss} = 10$, $20$ and $40 \, \hcMpc$ (corresponding to $2.11$, $4.21$ and $8.43$~pMpc, respectively).

We present two-dimensional slices through the simulation box in \fig{fig:Gamma_HI_cell}. The top row shows the ionization field, while the middle and bottom rows display the photoionization rate $\Gamma_{\mathrm{HI}, i}$ computed using the ray-tracing method and SCRIPT, respectively, for three different values of the mean free path $\lambda_\mathrm{ss}$.

As seen in the figure, our SCRIPT-based method produces noticeably smoother photoionization rate fields compared to the ray-tracing approach. This difference arises primarily because the ray-tracing method explicitly blocks ionizing radiation from propagating between points when the line of sight intersects a predominantly neutral cell. In contrast, SCRIPT's spherical averaging distributes the ionizing flux isotropically, regardless of direction, and thus cannot capture such anisotropic absorption. Additionally, we observe that the level of smoothing in the SCRIPT-computed $\Gamma_{\mathrm{HI}, i}$ field increases with increasing $\lambda_\mathrm{ss}$, thereby amplifying the discrepancy with the ray-tracing results at higher mean free paths.

To quantify this comparison, we plot the probability distribution function (PDF) of $\Gamma_{\mathrm{HI}, i}$ within ionized regions in \fig{fig:Gamma_HI_hist}. Vertical dashed lines indicate the mean photoionization rate in these regions for each method. The figure demonstrates that the agreement between the two methods is excellent—both qualitatively and quantitatively—for small values of $\lambda_\mathrm{ss}$, and gradually deteriorates as $\lambda_\mathrm{ss}$ increases, though the discrepancy remains within reasonable bounds.

This trend can be understood as follows: when the mean free path is small relative to the typical distance between ionizing sources and neutral absorbers, the contribution of radiation that would be absorbed by intervening neutral cells is minimal. Consequently, the omission of directional absorption in SCRIPT has a limited impact, and the resulting $\Gamma_{\mathrm{HI}, i}$ values closely match those from ray-tracing. Conversely, for larger values of $\lambda_\mathrm{ss}$, ionizing photons can travel farther, increasing the likelihood of intersecting neutral regions. In this case, the ray-tracing method accounts for significant absorption along certain sightlines, which SCRIPT, due to its isotropic treatment, fails to capture, leading to a disagreement in $\Gamma_{\mathrm{HI}, i}$. Indeed, we find that SCRIPT underestimates the mean photoionization rate in ionized regions by approximately 6\%, 11\%, and 19\% for $\lambda_\mathrm{ss} = 10$, $20$, and $40\,\hcMpc$, respectively. 

It is worth emphasizing, however, that observational constraints indicate a mean free path of $\approx 4~\hcMpc$ at $z \approx 6$. In this regime, SCRIPT performs very well in reproducing the ray-tracing results, indicating that our method is sufficiently accurate for the relevant astrophysical conditions. At lower redshifts, although the mean free path increases, the mass-averaged ionized fraction $Q_\mathrm{HII}$ also becomes larger. As a result, the impact of residual neutral islands—and hence the significance of directionally resolved absorption—diminishes. For the scenarios considered in this work, we therefore conclude that our method implemented in SCRIPT offers a computationally efficient yet physically reasonable approximation to more expensive ray-tracing calculations.

\section{Effect of Case B Recombination Coefficient}
\label{app:case_B}

\begin{figure}
\centering
\includegraphics[width=0.99\textwidth]{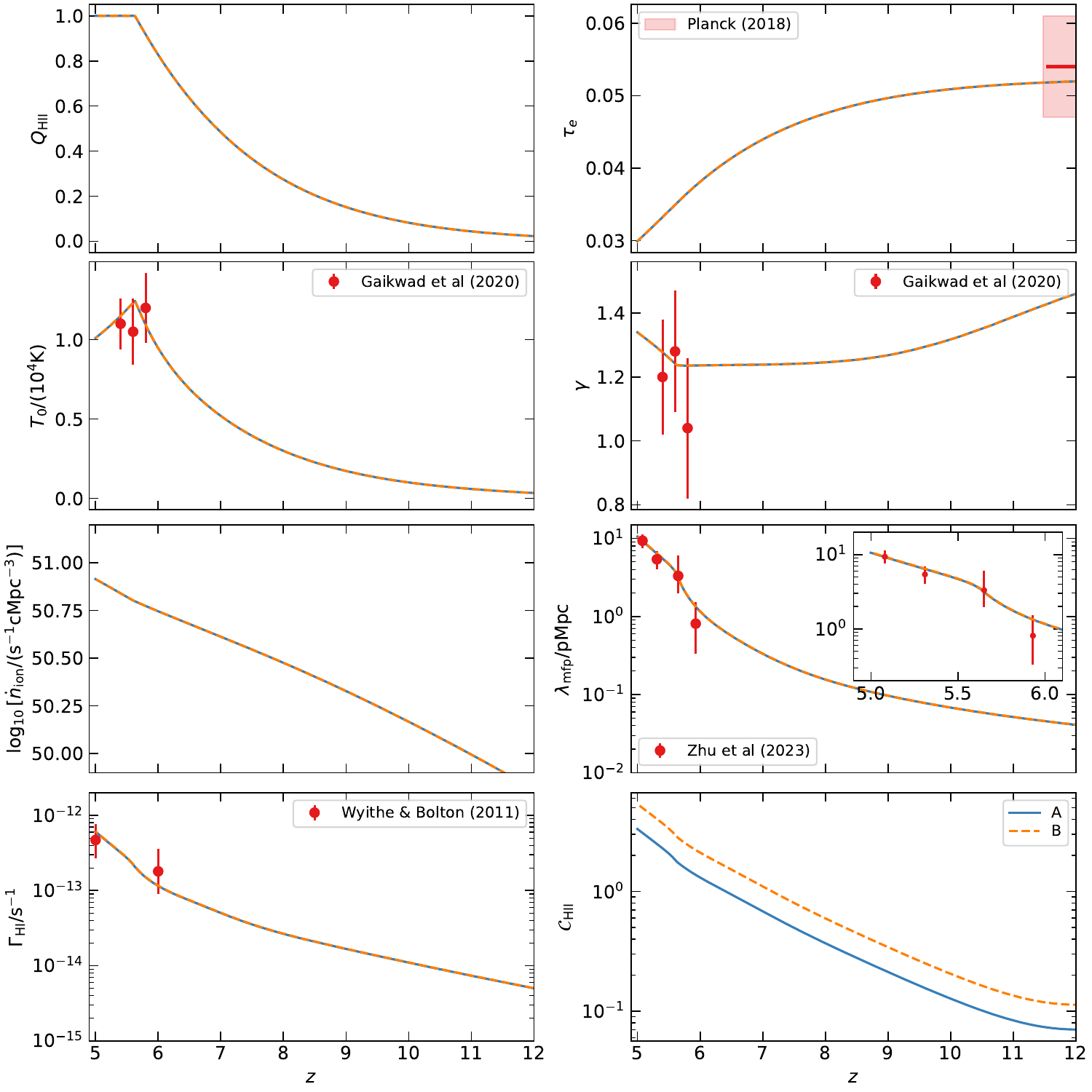}
\caption{Impact of switching from Case A to Case B recombination coefficient on various model outputs. All panels show results from the fiducial model using the default Case A (solid curves) and Case B (dashed curves) recombination, with the normalization $\mathcal{N}_{V,0}$ rescaled by a factor of $\approx 1.38$ to preserve the ionization and thermal histories. The top, middle and bottom left panels show that key observables such as the ionized and thermal histories, mean free path and photoionization rate remain unchanged. The bottom right panel shows the globally averaged clumping factor $\mathcal{C}_\mathrm{HII}$, which scales consistently by a factor of $\mathcal{R}_{A/B} \approx 1.6$ across redshift. This confirms that the recombination coefficient change can be fully absorbed by an appropriate rescaling of $\mathcal{N}_{V,0}$, except for an overall shift in $\mathcal{C}_\mathrm{HII}$.}
\label{fig:rec_variations}
\end{figure}

In this appendix, we assess the impact of adopting the Case B recombination coefficient in place of the default Case A on our results. To isolate the effect of this change, we consider how to adjust the model's free parameters so that key observables, such as the ionization and thermal histories, photoionization rate, and mean free path, remain unchanged under the transformation $\alpha_A \rightarrow \alpha_B$.

Since the temperature dependence of $\alpha_A(T)$ and $\alpha_B(T)$ is identical over the relevant temperature range, the difference between the two cases can be encapsulated by a constant scaling factor
\be
\mathcal{R}_{A/B} \equiv \f{\alpha_A(T)}{\alpha_B(T)} \approx 1.6.
\ee

The primary dependence on the recombination rate in our model enters through the product $C_{H,i} \, \alpha_A$ in \eqns{eq:dnrec_dt}{eq:epsilon_PH}. To maintain the same effective recombination rate when switching to Case B, the clumping factor must be rescaled as:
\be
C_{H,i} \rightarrow C_{H,i} ~ \mathcal{R}_{A/B}.
\ee

Recombination also influences the self-shielding threshold density $\Delta_{\mathrm{ss}, i}$, introduced in \secn{sec:clumping}, which scales as
\be
\Delta_{\mathrm{ss}, i} \propto \alpha_A^{-2/3}.
\ee
This, in turn, affects the clumping factor via its dependence on the density distribution:
\be
C_{H,i} \propto \mathcal{N}_{V,0} ~ \Delta_{\mathrm{ss}, i}^{3 - \beta_V} \propto \mathcal{N}_{V,0} ~ \alpha_A^{2\beta_V/3 - 2}.
\ee
Thus, to preserve the clumping factor’s scaling under $\alpha_A \rightarrow \alpha_B$, we must also rescale the normalization $\mathcal{N}_{V,0}$ of the density distribution as:
\be
\mathcal{N}_{V,0} \rightarrow \mathcal{N}_{V,0} ~ \mathcal{R}_{A/B}^{2\beta_V/3 - 1}.
\ee
With this transformation, the resulting clumping factor scales precisely as required to compensate for the change in $\alpha_A$, leaving all other model outputs invariant.

We validate this analytically motivated scaling by explicitly implementing the Case B recombination rate in our code, while simultaneously rescaling $\mathcal{N}_{V,0} \rightarrow 1.38 \, \mathcal{N}_{V,0}$ for the fiducial value of $\beta_V = 2.52$. The results are shown in \fig{fig:rec_variations}. As expected, all derived quantities, except the globally averaged clumping factor $\mathcal{C}_\mathrm{HII}$, remain unchanged. The clumping factor itself scales uniformly with redshift by a factor of $\mathcal{R}_{A/B} \approx 1.6$, as shown in the bottom right panel of the figure.

Therefore, when using Case B recombination instead of Case A, all estimates of the ionized hydrogen clumping factor $\mathcal{C}_\mathrm{HII}$ in this work should be multiplied by a factor of 1.6 to remain consistent with the original model predictions.

\section{Convergence with respect to Grid Size}
\label{app:resolution}

\begin{figure}
    \centering
    \includegraphics[width=0.99\textwidth]{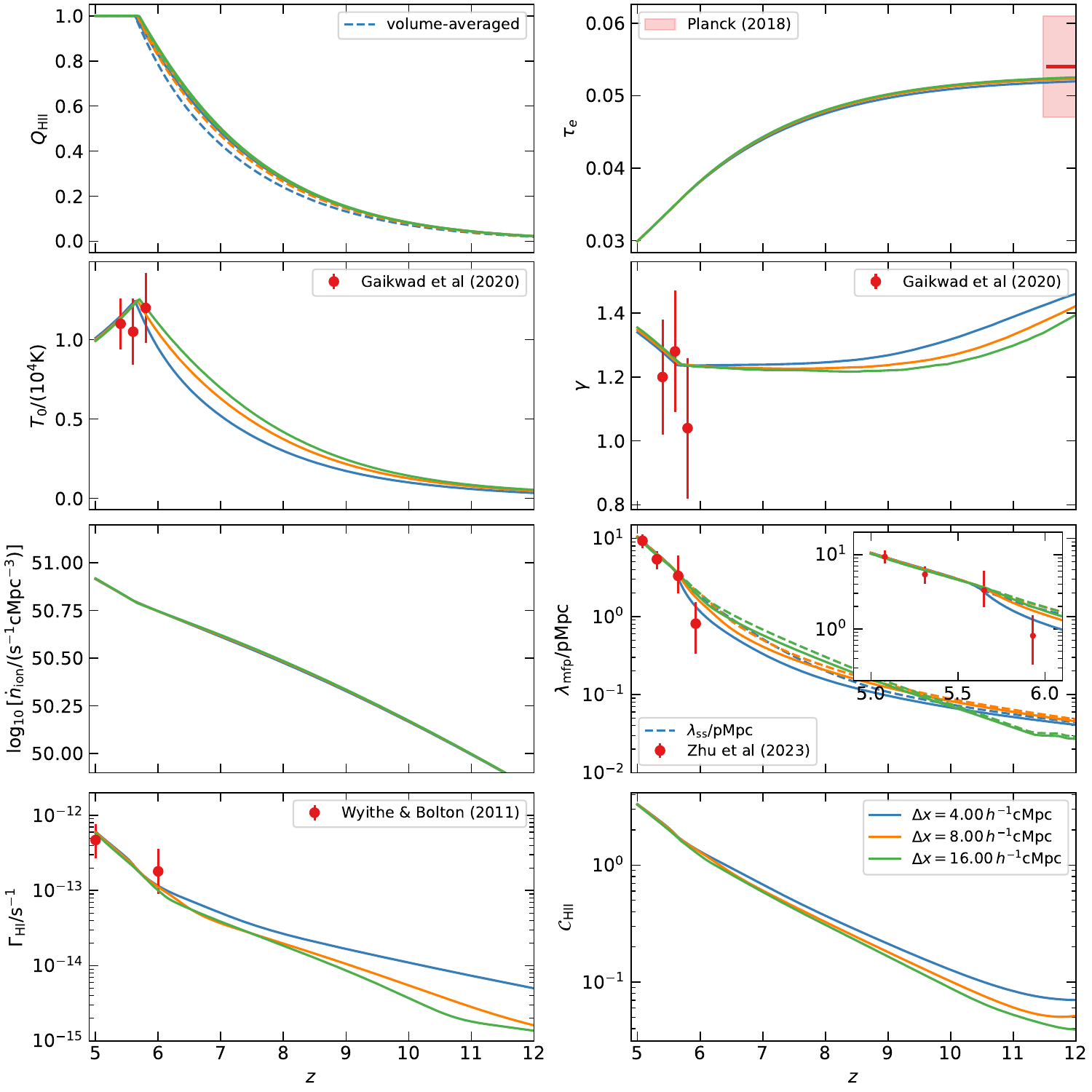}
    \caption{Dependence of the evolution of globally-averaged physical quantities on the resolution of the simulation. The simulation box has been kept the same, except the grid size for generating the ionization, thermal and other fields which has been varied. While changing the resolution, we keep all the model parameters unchanged, except the ones related to the conditional density distribution, which depends on the grid cell size. In practice, however, only one parameter $\gamma_V$ is required to be modified with resolution. It is clear that the ionizing emissivity is independent of resolution (which is a consequence of the luminosity function being unaffected by grid size, not shown here), which leads to resolution-independent evolution of the mass-averaged ionized fraction. We also see that the IGM parameters are affected by the grid resolution at high redshifts, mainly because of the inability of our sub-grid model to account for correlation between the density and ionization fields below the grid scale. This is clear from the non-convergence of the volume-averaged ionized fraction, shown by dashed lines in the top left panel. It is also to be noted that the non-convergence of quantities like the mean free path $\lambda_\mathrm{mfp}$ (third row right panel) and photoionization rate $\Gamma_\mathrm{HI}$ (bottom row left panel) are less than the typical observational errors at high redshifts.}
    \label{fig:resolution_test}
\end{figure}

We now examine the dependence of our results on the size of the simulation grid cells. Our default grid size is $\Delta x = 4 \hcMpc$, and we compare the results with two coarser resolutions, $\Delta x = 8$ and $16 \hcMpc$. When analyzing resolution dependence, it is crucial to account for the fact that the conditional density distribution $P_V(\Delta | \Delta_i)$ is defined in terms of the grid size used to compute the density $\Delta_i$. As a result, the parameters $\mathcal{N}_{V,0}$, $\alpha_V$, $\gamma_V$, and $\beta_V$ that define the conditional PDF are naturally resolution-dependent.

For the default resolution, we assume that all grid cells, as well as the unconditional PDF $P_V(\Delta)$, share the same value of $\beta_V$, ensuring that the high-density tail of the PDF has an identical shape across all regions. Since the slope of the unconditional distribution cannot depend on the grid size, $\beta_V$ must be resolution-independent.

\begin{table}
    \centering
    \begin{tabular}{|c|c|}
        \hline
        $\Delta x / (\hcMpc)$ & $\gamma_V$ \\ \hline
        $4.0$                 & $-0.02$    \\ \hline
        $8.0$                 & $0.70$     \\ \hline
        $16.0$                & $2.11$     \\ \hline
    \end{tabular}
    \caption{Dependence of the parameter $\gamma_V$ on the grid size used to compute the ionization and thermal histories. Since this parameter characterizes the conditional density PDF $P_V(\Delta | \Delta_i)$, it is expected that it would depend on the grid size at which the large-scale density field $\Delta_i$ is computed. The $\gamma_V$ values are chosen so as to ensure the unconditional density distribution $P(\Delta)$ to remain resolution-independent.}
    \label{tab:gamma_v_res}
\end{table}

To determine the resolution dependence of the remaining parameters ($\mathcal{N}_{V,0}$, $\alpha_V$, and $\gamma_V$), we enforce the condition that the amplitude $\left \langle \mathcal{N}_V(\Delta_i) \right \rangle = P_V(\Delta) \, \Delta^{\beta_V}$ remains resolution-independent at all redshifts. For coarser resolutions, we adjust $\mathcal{N}_{V,0}$, $\alpha_V$, and $\gamma_V$ to satisfy this condition, finding that only $\gamma_V$ needs to vary with grid size. Consequently, while testing convergence, we fix all other parameters to their fiducial values and vary only $\gamma_V$ to ensure that $\left \langle \mathcal{N}_V(\Delta_i) \right \rangle$ remains unchanged. The values of $\gamma_V$ for different grid resolutions are shown in \tab{tab:gamma_v_res}. As the grid cell size $\Delta x$ decreases, the value of $\gamma_V$ also decreases. This change reduces the normalization of the density PDF in overdense cells while increasing it in underdense cells, see \eqn{eq:N_V_Delta_i}. In effect, a smaller cell size leads to reduced fluctuations in the density field. This result is consistent with the idea that averaging the density over a region of size $\Delta x$ removes contributions from larger-scale fluctuations, leaving primarily the smaller-scale ones. In other words, when $\Delta x$ is decreased, more of the total fluctuation power is captured by the large-scale density, resulting in fewer fluctuations on smaller scales. This behavior aligns with the characteristics of the conditional density distribution observed in cosmological Gaussian random fields.

The evolution of globally-averaged quantities with resolution is shown in \fig{fig:resolution_test}. The figure demonstrates that the emissivity and reionization history are well-converged with respect to resolution. Additionally, thermal parameters $T_0$ and $\gamma$, as well as quantities like $\Gamma_\mathrm{HI}$, $\lambda_\mathrm{mfp}$, and $\mathcal{C}_\mathrm{HII}$, are converged in the post-reionization era. However, some resolution dependence is observed when the universe is partially ionized. Importantly, this resolution dependence is significantly smaller than the uncertainties in the corresponding observational measurements.

The observed resolution dependence can be attributed to the inability of coarse grids to fully capture the correlation between the matter density and ionized fraction below the grid scale. In particular, the volume-averaged ionized fraction in partially ionized cells is not fully converged \emph{for any simulation of reionization}, unless the resolution is fine enough to elimnate the partially ionized cells. For example, in an inside-out reionization scenario, ionized regions are preferentially concentrated in high-density regions, resulting in a smaller volume coverage compared to the ionized mass fraction. Similarly, the volume coverage of neutral regions would exceed the mass-averaged neutral fraction. This discrepancy implies that the fraction of equal-volume points residing in neutral regions within a grid cell is larger than $1 - x_{\mathrm{HII}, i}$. As a result, coarse resolutions can underestimate the contribution of neutral regions when computing the mean free path, leading to an overestimation of $\lambda_\mathrm{mfp}$ for coarser grids, as seen in \fig{fig:resolution_test}. Similar effects are observed for other quantities as well.

To achieve convergence in the volume-averaged ionized fraction, one would need to use grid sizes small enough to eliminate partially ionized cells, which would entail significant computational costs. Alternatively, this aspect could be modeled using sub-grid physics in a semi-analytical framework. However, given that the lack of convergence is smaller than observational uncertainties, we defer this study to future work.

\section{Convergence with respect to Simulation Volume}
\label{app:box_size}

\begin{figure}
    \centering
    \includegraphics[width=0.99\textwidth]{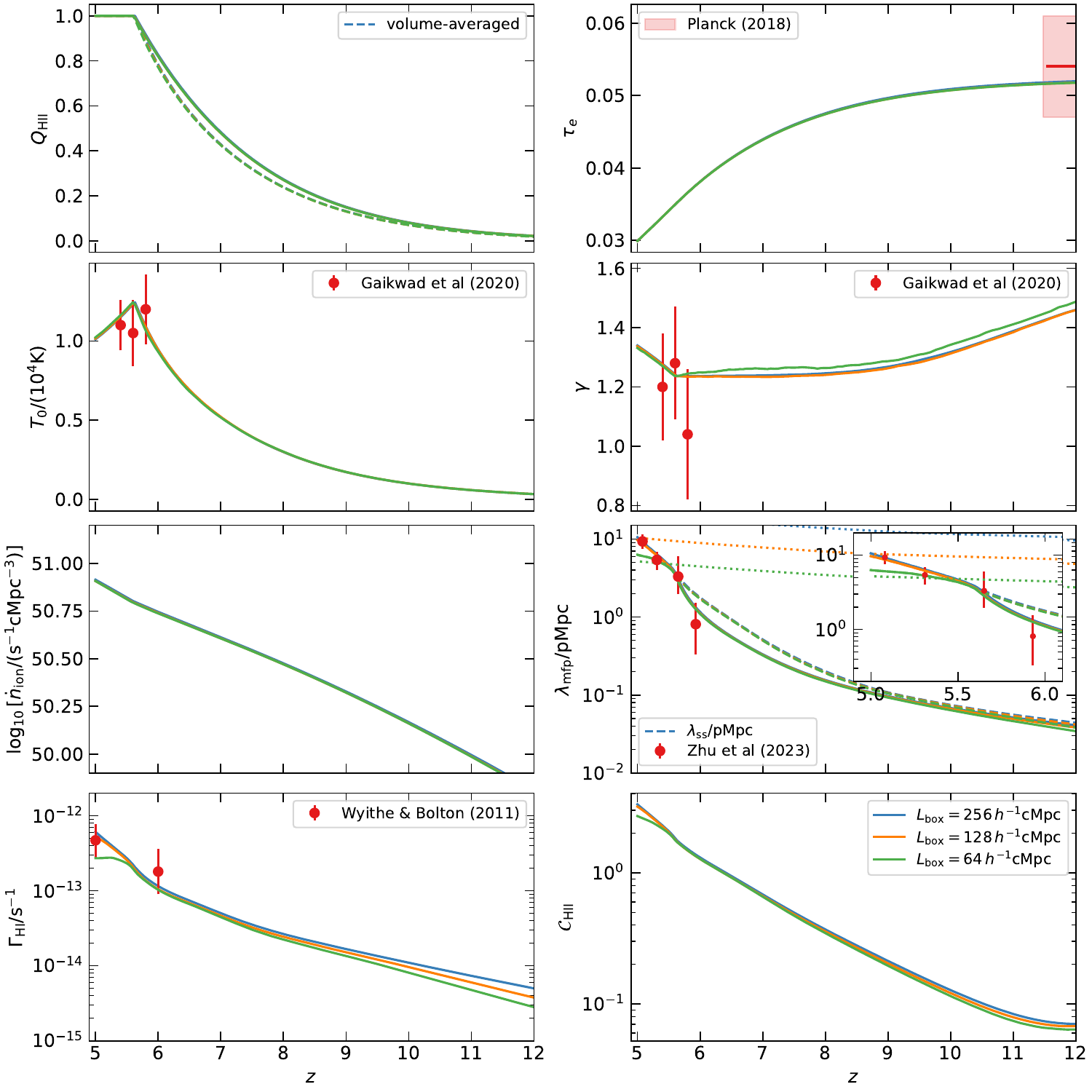}
    \caption{Dependence of the evolution of globally-averaged physical quantities on the simulation volume. The grid cell size as well as all the model parameters are kept the same while comparing results for different box sizes. It is clear that the size of the box does not affect our results in any significant way. Its only effect is to underestimate the mean free path $\lambda_\mathrm{mfp}$ (third row right panel) when it becomes about one-third the box length (indicated using dotted lines). This effect propagates to other quantities like the photoionization rate $\Gamma_\mathrm{HI}$ (bottom row left panel) and the clumping factor $\mathcal{C}_\mathrm{HII}$ (bottom row right panel). Our default box size is sufficient to model the mean free path as given by the observational data.}
    \label{fig:boxsize_test}
\end{figure}

We now examine the convergence of our results with respect to the simulation volume, or box size. Our default simulation box has a length of $L_\mathrm{box} = 256 \hcMpc$, and we compare the results with two smaller boxes of lengths $L_\mathrm{box} = 128 \hcMpc$ and $L_\mathrm{box} = 64 \hcMpc$. To ensure a fair comparison, the grid size is fixed at $\Delta x = 4 \hcMpc$ for all cases. The results of this analysis are presented in \fig{fig:boxsize_test}.

We find that almost all quantities are converged with respect to the simulation volume. However, deviations appear for $\lambda_\mathrm{mfp}$ in the smallest box ($L_\mathrm{box} = 64 \hcMpc$) at low redshifts. These deviations also affect related quantities, such as $\Gamma_\mathrm{HI}$ and $\mathcal{C}_\mathrm{HII}$. This behavior arises from the way photons are distributed within the simulation volume. The distance from a source is computed using periodic boundary conditions, and once this distance exceeds half the box size, photon propagation is halted. This prevents the same cell from ``seeing'' the same source multiple times due to periodicity. Any remaining photons are then redistributed uniformly across all cells, effectively representing an isotropic background radiation field. This approach also improves computational efficiency, since otherwise photons could be tracked to arbitrarily large distances, significantly increasing runtime. When the mean free path, $\lambda_\mathrm{mfp}$, is much smaller than the box size, this procedure has negligible impact, as the photon flux at a distance of half the box size is already minimal. However, when $\lambda_\mathrm{mfp}$ approaches the box size, the method can lead to an underestimation of $\lambda_\mathrm{mfp}$.

To illustrate this effect, we indicate $L_\mathrm{box} / 3$ using dotted lines in the mean free path panel of \fig{fig:boxsize_test}. The figure clearly shows that $\lambda_\mathrm{mfp}$ starts to be underestimated when it approaches one-third of the box length. For $L_\mathrm{box} = 128 \hcMpc$, the box size is barely sufficient to account for the data point at $z \sim 5$. In contrast, our default box ($L_\mathrm{box} = 256 \hcMpc$) provides sufficient volume to accommodate the largest mean free paths relevant to this work.

This analysis highlights the importance of choosing an appropriate box size for accurately modeling the mean free path and related quantities. The results demonstrate that the default box size used in our simulations is adequate for the redshift range and physical quantities considered in this study.

\bibliographystyle{JHEP}
\bibliography{main} 

\end{document}